\documentclass{article}

\usepackage{PRIMEarxiv}

\usepackage[utf8]{inputenc} 
\usepackage[T1]{fontenc}    
\usepackage{hyperref}       
\usepackage{url}            
\usepackage{booktabs}       
\usepackage{amsfonts}       
\usepackage{nicefrac}       
\usepackage{microtype}      
\usepackage{lipsum}
\usepackage{fancyhdr}       
\usepackage{graphicx}       
\graphicspath{{media/}}     
\usepackage{authblk}
\usepackage{amsmath}
\usepackage{bm}
\usepackage{algorithm}
\usepackage{algorithmic}
\usepackage{float}       
\usepackage{booktabs}    
\usepackage{tabularx}    
\usepackage{multirow,booktabs}
\newcolumntype{C}{>{\centering\arraybackslash}X}

\begin{document}
\title{Spatiotemporal Graph Contrastive Learning with Dual-Graph Regularized Sparse Canonical Correlation Analysis for Longitudinal Imaging–Multi-Omics Integration in Early Mild Cognitive Impairment}
\author[1]{Jin Deng}
\author[1]{Tao Xu}
\author[1]{Yinning Luo}
\author[1]{Mulian Wang}
\author[2]{Ying Li}
\author[3]{Nizhuan Wang\textsuperscript{\kern0.25em *,}}

\affil[1]{College of Mathematics and Informatics, South China Agricultural University, Guangzhou 510642, China}
\affil[2]{Faculty of Intelligence Technology, Shanghai Institute of Technology, Shanghai 201418, China}
\affil[3]{Department of Language Science and Technology, The Hong Kong Polytechnic University, Hung Hom, Kowloon, Hong Kong SAR, China}

\maketitle
\renewcommand{\thefootnote}{*}
\footnotetext{Corresponding author. Email: wangnizhuan1120@gmail.com}
\renewcommand{\thefootnote}{\arabic{footnote}}

\begin{abstract}
Early mild cognitive impairment (EMCI) is a critical neurobiological window for blocking the conversion from Alzheimer's disease (AD) to dementia. However, longitudinal brain imaging is affected by measurement noise and nonlinear atrophy. Existing longitudinal association algorithms are limited by single-time-point slices and linear temporal assumptions. They also lack mechanisms for integrating epigenetics and molecular intrinsic topology modeling. To address this, this study proposes a two-stage framework characterized by deep spatiotemporal purification followed by topology-cooperative target hunting. The front end builds a spatiotemporal graph contrastive learning feature purification engine (ST-GCL). It uses graph convolution and dynamic temporal self-attention. In an unsupervised manner, it purifies multi-time-point longitudinal brain imaging sequences into a single, highly robust individual-level spatiotemporal representation. The back end builds a dual-graph regularized multi-task SCCA (DG-MTSCCA) model. With molecular topology priors, it enables dual-omics cooperative target hunting between dynamic imaging evolution trajectories and underlying SNPs and DNA methylation. Simulation tests and experiments on the real ADNI longitudinal cohort show that this framework outperforms existing advanced longitudinal and multimodal models overall. It achieves this in association mapping accuracy, feature detection rate under high noise and small samples, and downstream EMCI early classification and diagnostic performance. In addition, longitudinal association consistency analysis not only identifies established AD risk genes and vulnerable cortical networks, but also highlights a set of exploratory candidate biomarkers consistently associated with longitudinal brain structural trajectories. This study bridges the scale gap between brain spatiotemporal phenotypes and omics association analysis. It provides a new computational system for in-depth analysis of the early conversion mechanisms of Alzheimer's disease.The source code for the proposed STGSCCA framework is publicly available at 
\url{https://github.com/XT2024-OO/STGSCCA}.
\end{abstract}


\section{Introduction}
Alzheimer's disease (AD) is an insidious, irreversible, and progressive neurodegenerative disease. It has become one of the leading causes of death and disability among the elderly worldwide \cite{ref2,ref3}. Repeated failures in clinical trials indicate that once AD progresses to the dementia stage, brain tissue damage is already irreversible. Therefore, research focus is inevitably shifting toward the early stages of the disease. Early mild cognitive impairment (EMCI), as a critical early stage in the transition from normal aging to AD dementia, is the best ``time window'' for exploring pathological mechanisms and conducting clinical intervention \cite{ref4,ref5}. In the EMCI stage, clinical symptoms are often extremely subtle, but the underlying molecular networks and local brain anatomical structures have already begun silent pathological cascades. Therefore, deeply analyzing the disease evolution patterns at the EMCI stage has crucial clinical value for delaying or even preventing the onset of AD \cite{ref6}.

To understand EMCI, we need to link changes in brain structure with changes at the molecular level. Longitudinal structural magnetic resonance imaging (longitudinal sMRI) can intuitively and dynamically capture the macroscopic cortical atrophy trajectory caused by the disease along the temporal axis \cite{ref7,ref8}. At the microscopic level, single nucleotide polymorphisms (SNPs) provide the underlying genetic baseline that determines individual susceptibility \cite{ref9}, while DNA methylation characterizes environmentally driven dynamic epigenetic modifications \cite{ref10}. Conducting multimodal association studies across these complementary data sources aims to delineate the biological cascade from genotypes and epigenetic modifications to macroscopic brain phenotypes, thereby uncovering key molecular factors underlying progressive brain atrophy. However, jointly analyzing such longitudinal high-dimensional data faces enormous challenges. Longitudinal sMRI sequences are accompanied not only by high-frequency measurement noise introduced by device calibration and physiological fluctuations, but brain atrophy itself has highly nonlinear dynamic temporal characteristics and individual heterogeneity. Meanwhile, SNP and DNA methylation data also reside in extremely high-dimensional omics spaces, containing complex genetic background noise \cite{ref11,ref12}. If these spatiotemporal noise from multiple sources cannot be effectively filtered during analysis, the extremely weak early pathological signals of EMCI will be completely overwhelmed, causing the model to fit numerous biologically meaningless spurious correlations \cite{ref13}.

\begin{figure*}[h]   
\centering
\includegraphics[width=1\linewidth]{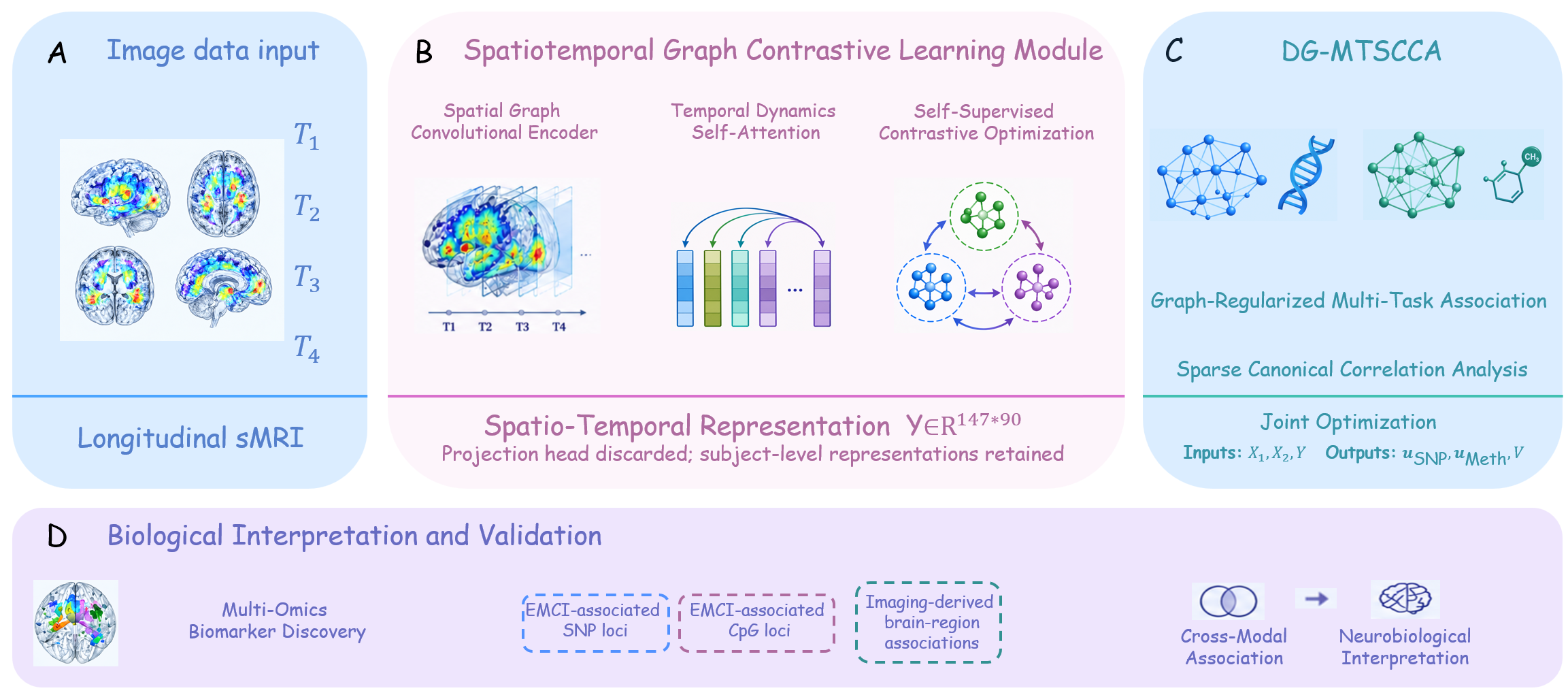}   
\caption{Overall flowchart of the proposed STGSCCA framework. (A) Longitudinal sMRI data input across four time points. (B) Spatiotemporal graph contrastive learning (ST-GCL) module for spatial denoising and temporal trajectory extraction. (C) Dual-graph multi-task sparse canonical correlation analysis (DG-MTSCCA) for joint multi-omics and imaging association. (D) Biological interpretation and clinical validation. }
\label{fig_1}
\end{figure*}

In the field of imaging genetics, sparse canonical correlation analysis (SCCA) is a classic algorithm for discovering multivariate associations between omics and imaging data \cite{ref14,ref15,ref16}. Early SCCA models were mainly designed for cross-sectional (single-time-point) data and could not capture disease evolution trends \cite{ref17}. In recent years, to adapt to the characteristics of longitudinal data, researchers have made a series of longitudinal extensions based on traditional SCCA. For example, by introducing regularization terms such as temporal fused lasso or group sparsity, imaging data from different time points are treated as continuous sequences or multiple tasks, forcing the projection weights of adjacent time points to remain smooth or consistent. For example, Hao et al.\ introduced group sparsity and fusion penalty terms into the model to force feature weights at adjacent time points to remain smooth, thereby breaking through the limitation of single-time-point analysis and successfully capturing the longitudinal evolution patterns of brain imaging features associated with DNA methylation variation over time \cite{ref18}. Du et al.\ innovatively proposed the multi-task temporal sparse canonical correlation analysis (TMTSCCA) framework, treating longitudinal analysis at different follow-up time points as a multi-task joint learning process \cite{ref19}. Deng et al.\ proposed a spatiotemporal sparse canonical correlation analysis framework. By simultaneously introducing temporal smoothness constraints and spatial structure penalty terms into the longitudinal model, it successfully filtered out temporal noise and precisely captured the complex longitudinal evolution patterns of brain imaging features driven by DNA methylation variation over time \cite{ref21}. Although these longitudinal temporal SCCA models have improved association accuracy to some extent, they essentially still rely on linear mathematical constraints to handle temporal evolution. Faced with highly nonlinear brain atrophy trajectories in real clinical sequences and complex spatially localized measurement noise, traditional mathematical models lack the ability to perform deep dimensionality reduction and spatiotemporal feature purification on high-dimensional data before association analysis. This severely limits their target discovery accuracy in EMCI cohorts with extreme noise and weak signals.
To address the limitations of traditional longitudinal SCCA models in nonlinear temporal modeling and noise resistance, this study proposes a novel two-stage framework termed Spatiotemporal Graph Sparse Canonical Correlation Analysis (STGSCCA), whose overall framework is shown in Figure~\ref{fig_1}. Specifically, STGSCCA integrates a front-end spatiotemporal graph contrastive learning module (ST-GCL) with a back-end dual-graph regularized multi-task SCCA model (DG-MTSCCA). Based on longitudinal sMRI sequences across four follow-up time points paired with baseline DNA methylation and SNP data in the ADNI cohort, this study aims to strip longitudinal clinical heterogeneous noise through the synergy of front-end spatiotemporal feature purification and back-end mathematical optimization, thereby robustly and efficiently mining core microscopic multi-omics targets and macroscopic imaging biomarkers associated with cognitive decline at the EMCI stage. The main contributions of this paper are summarized as follows:

\begin{enumerate}
     \item Spatiotemporal graph contrastive learning purification engine (ST-GCL) is proposed for the first time. High-noise multi-time-point longitudinal brain imaging sequences are unsupervisedly compressed and purified into a single spatiotemporal representation that integrates nonlinear disease evolution patterns.
    
    \item A dual-graph multi-task framework is constructed by integrating front-end deep spatiotemporal purification with back-end topology-cooperative association, enabling precise cross-modal target identification linking dynamic brain phenotypes with underlying SNPs and DNA methylation.
    
    \item Multi-omics longitudinal association consistency analysis is conducted by linking structural MRI, DNA methylation, and SNPs at the EMCI stage. Beyond confirming canonical risk genes, exploratory candidate biomarkers with high discriminative capability are uncovered.
\end{enumerate}

\section{Methods}

\subsection{Front End: Spatiotemporal Graph Contrastive Learning Module }
Longitudinal structural magnetic resonance imaging (sMRI) is susceptible to local measurement noise introduced by device scanning and postural artifacts during acquisition. Meanwhile, the brain atrophy process caused by neurodegeneration shows significant nonlinear evolution characteristics along the temporal dimension. To suppress spatial measurement noise and characterize the longitudinal developmental trajectory of the disease under unsupervised conditions, this study constructs a spatiotemporal graph contrastive learning module (ST-GCL). This module contains three cascaded core components: graph convolutional encoding combined with spatial priors, dynamic temporal self-attention fusion, and longitudinal graph contrastive supervision based on InfoNCE. Assume that the longitudinal clinical cohort in this study contains N independent subjects, each with T follow-up time points, and the brain is divided into V regions of interest (ROIs) by an anatomical template.
\subsubsection{Spatial Graph Convolutional Encoder}
A single MRI scan is susceptible to machine noise and micromotion artifacts, causing local brain region measurements to deviate from their true values. To use morphological covariance patterns among brain regions for spatial denoising, this module models the brain as an undirected graph \(\mathbf{G}=(\mathbf{V},\mathbf{E})\). The node set \(\mathbf{V}\) represents \(\mathbf{V}\) independent brain regions, and the edge set \(\mathbf{E}\) represents morphological similarity associations among brain regions. A prior adjacency matrix \(\mathbf{A}\in\mathbb{R}^{\mathbf{V}\times \mathbf{V}}\) is thus constructed.

To strictly follow the cross-validation principle and avoid cross-sample data leakage, \(\mathbf{A}\) is constructed only on subjects in the independently partitioned training set based on a morphological similarity network (MSN). Let the brain region feature matrix of the training set at time point \(\mathbf{t}\) be \(\mathbf{M}\in\mathbb{R}^{\mathbf{N}_{\mathrm{train}}\times \mathbf{V}}\). The Pearson correlation coefficient between brain region \(\mathbf{j}\) and brain region \(\mathbf{k}\) is defined as:
\begin{equation}
\mathbf{S}_{\mathbf{jk}}
=
\frac{
\sum_{\mathbf{m}=1}^{\mathbf{N}_{\mathrm{train}}}
\left(\mathbf{M}_{\mathbf{m},\mathbf{j}}-\bar{\mathbf{M}}_{\mathbf{j}}\right)
\left(\mathbf{M}_{\mathbf{m},\mathbf{k}}-\bar{\mathbf{M}}_{\mathbf{k}}\right)
}{
\sqrt{
\sum_{\mathbf{m}=1}^{\mathbf{N}_{\mathrm{train}}}
\left(\mathbf{M}_{\mathbf{m},\mathbf{j}}-\bar{\mathbf{M}}_{\mathbf{j}}\right)^{2}
}
\sqrt{
\sum_{\mathbf{m}=1}^{\mathbf{N}_{\mathrm{train}}}
\left(\mathbf{M}_{\mathbf{m},\mathbf{k}}-\bar{\mathbf{M}}_{\mathbf{k}}\right)^{2}
}
}
\end{equation}
Here, \(\bar{\mathbf{M}}_{\mathbf{j}}\) denotes the mean of brain region \(\mathbf{j}\) in the training cohort. Subsequently, the similarity matrix \(\mathbf{S}\) is binarized and truncated using a preset significance threshold \(\boldsymbol{\theta}\) to construct the adjacency matrix \(\mathbf{A}\):
\begin{equation}
\mathbf{A}_{\mathbf{jk}}=
\begin{cases}
\mathbf{1}, & \text{if } \mathbf{j}\neq \mathbf{k} \text{ and } |\mathbf{S}_{\mathbf{jk}}|\geq \boldsymbol{\theta},\\
\mathbf{0}, & \text{otherwise}.
\end{cases}
\end{equation}

Let \(\mathbf{X}_{\mathbf{i},\mathbf{t}}\in\mathbb{R}^{\mathbf{V}\times \mathbf{d}}\) be the initial brain imaging feature input matrix of subject \(\mathbf{i}\) at time point \(\mathbf{t}\) (\(\mathbf{i}\in\{\mathbf{1},\mathbf{2},\dots,\mathbf{N}\}\), \(\mathbf{t}\in\{\mathbf{1},\mathbf{2},\dots,\mathbf{T}\}\), and \(\mathbf{d}\) is the initial feature dimension of a single brain region). This module uses a multi-layer graph convolutional network (GCN) for spatial smoothing and feature aggregation \cite{ref22}. The feature propagation formula of its \(\mathbf{l}\)-th layer is defined as follows:
\begin{equation}
\mathbf{H}_{\mathbf{i},\mathbf{t}}^{(\mathbf{l}+\mathbf{1})}
=
\boldsymbol{\sigma}\!\left(
\tilde{\mathbf{D}}^{-\mathbf{1}/\mathbf{2}}
\tilde{\mathbf{A}}
\tilde{\mathbf{D}}^{-\mathbf{1}/\mathbf{2}}
\mathbf{H}_{\mathbf{i},\mathbf{t}}^{(\mathbf{l})}
\mathbf{W}^{(\mathbf{l})}
\right)
\end{equation}
The physical and mathematical meanings of each symbol are as follows:

where \(\mathbf{H}_{\mathbf{i},\mathbf{t}}^{(\mathbf{l})}\) and \(\mathbf{H}_{\mathbf{i},\mathbf{t}}^{(\mathbf{l}+\mathbf{1})}\) represent the hidden layer feature matrices of subject \(\mathbf{i}\) at time point \(\mathbf{t}\) after the \(\mathbf{l}\)-th and \((\mathbf{l}+\mathbf{1})\)-th layers of network convolution, respectively (when \(\mathbf{l}=\mathbf{0}\), \(\mathbf{H}_{\mathbf{i},\mathbf{t}}^{(\mathbf{0})}=\mathbf{X}_{\mathbf{i},\mathbf{t}}\)). \(\tilde{\mathbf{A}}=\mathbf{A}+\mathbf{I}\) is the self-connected adjacency matrix with the identity matrix \(\mathbf{I}\) (self-loops) introduced, ensuring that brain regions retain their own basic features when aggregating neighbor information. \(\tilde{\mathbf{D}}\) is the degree matrix corresponding to matrix \(\tilde{\mathbf{A}}\), whose diagonal elements are defined as \(\tilde{\mathbf{D}}_{\mathbf{jj}}=\sum_{\mathbf{k}}\tilde{\mathbf{A}}_{\mathbf{jk}}\), and \(\tilde{\mathbf{D}}^{-\mathbf{1}/\mathbf{2}}\) is used to symmetrically normalize the adjacency matrix, preventing gradient explosion of features during multi-layer propagation. \(\mathbf{W}^{(\mathbf{l})}\) is the learnable weight parameter matrix of the \(\mathbf{l}\)-th layer graph convolutional network. \(\boldsymbol{\sigma}(\cdot)\) represents the ReLU (Rectified Linear Unit) nonlinear activation function.

After completing the spatial smoothing of multi-layer graph convolution, the model maps the multi-dimensional hidden layer features to scalar outputs through node-level linear projection, obtaining the denoised brain region feature vector at the current time point:
\begin{equation}
\mathbf{h}_{\mathbf{i},\mathbf{t}}
=
\mathbf{H}_{\mathbf{i},\mathbf{t}}^{(\mathbf{L})}
\mathbf{W}_{\mathrm{proj}}
\end{equation}
Here, \(\mathbf{h}_{\mathbf{i},\mathbf{t}}\in\mathbb{R}^{\mathbf{V}}\) represents the purified feature vector of subject \(\mathbf{i}\) at time point \(\mathbf{t}\) after spatial denoising, \(\mathbf{H}_{\mathbf{i},\mathbf{t}}^{(\mathbf{L})}\) represents the final hidden layer matrix output after all \(\mathbf{L}\) layers of graph convolution, and \(\mathbf{W}_{\mathrm{proj}}\) is the linear projection weight vector that compresses multi-dimensional features into a single value.
\subsubsection{Temporal Dynamics Self-Attention}
During the progression along the Alzheimer's disease continuum, neurodegenerative atrophy manifests at heterogeneous rates across follow-up intervals, where conventional static aggregation (such as uniform temporal averaging) risks obscuring critical stages of accelerated phenotypic change. Therefore, this module introduces an adaptive temporal attention pooling mechanism to selectively prioritize informative follow-up time points according to their disease-associated morphological relevance \cite{ref23}.

The spatially purified features of subject $i$ across $T$ follow-up time points are structured into a longitudinal feature sequence $\mathbf{E}_i = [\mathbf{h}_{i,1}, \mathbf{h}_{i,2}, \dots, \mathbf{h}_{i,T}] \in \mathbb{R}^{V \times T}$. To evaluate the relative informative contribution of each follow-up stage toward the overall disease profile, a feedforward attention network consisting of two linear transformations is employed to compute the unnormalized importance score $\mathbf{e_{i,t}}$ for each timepoint $\mathbf{t}$:
\begin{equation}
\mathbf{e}_{\mathbf{i},\mathbf{t}}
=
\mathbf{w}_{\mathbf{a}}^{\top}
\tanh\!\left(
\mathbf{W}_{\mathbf{h}}
\mathbf{h}_{\mathbf{i},\mathbf{t}}
+
\mathbf{b}_{\mathbf{h}}
\right)
\tag{5}
\end{equation}
where \(\mathbf{e}_{\mathbf{i},\mathbf{t}}\) represents the unnormalized attention score of subject \(\mathbf{i}\) at time point \(\mathbf{t}\), \(\mathbf{W}_{\mathbf{h}}\) is the weight matrix of the first layer of the attention network, \(\mathbf{b}_{\mathbf{h}}\) is the bias vector of the first layer of the attention network, \(\tanh(\cdot)\) is the hyperbolic tangent activation function that introduces nonlinear fitting capability, and \(\mathbf{w}_{\mathbf{a}}^{\top}\) is the transpose of the weight vector of the second layer of the attention network, used to map the hidden layer to a scalar score.

Subsequently, the Softmax function is used to globally normalize the scores of each time point along the longitudinal temporal axis, yielding the dynamic attention weight \(\boldsymbol{\alpha}_{\mathbf{i},\mathbf{t}}\):
\begin{equation}
\boldsymbol{\alpha}_{\mathbf{i},\mathbf{t}}
=
\frac{
\exp\!\left(\mathbf{e}_{\mathbf{i},\mathbf{t}}\right)
}{
\sum_{\mathbf{k}=\mathbf{1}}^{\mathbf{T}}
\exp\!\left(\mathbf{e}_{\mathbf{i},\mathbf{k}}\right)
}
\tag{6}
\end{equation}
where \(\boldsymbol{\alpha}_{\mathbf{i},\mathbf{t}}\) represents the relative importance weight assigned to time point \(\mathbf{t}\) (\(\mathbf{0}<\boldsymbol{\alpha}_{\mathbf{i},\mathbf{t}}<\mathbf{1}\) and \(\sum_{\mathbf{t}=\mathbf{1}}^{\mathbf{T}}\boldsymbol{\alpha}_{\mathbf{i},\mathbf{t}}=\mathbf{1}\)), \(\exp(\cdot)\) is the exponential function with the natural constant \(\mathbf{e}\) as the base, and \(\mathbf{k}\) is the traversal index of the time series, ranging from \(\mathbf{1}\) to \(\mathbf{T}\).

Finally, the model performs weighted summation of the multi-time-step spatially denoised features to compress the three-dimensional temporal sequence into a single individual-level spatiotemporal joint representation vector \(\mathbf{z}_{\mathbf{i}}\):
\begin{equation}
\mathbf{z}_{\mathbf{i}}
=
\sum_{\mathbf{t}=\mathbf{1}}^{\mathbf{T}}
\boldsymbol{\alpha}_{\mathbf{i},\mathbf{t}}
\mathbf{h}_{\mathbf{i},\mathbf{t}}
\tag{7}
\end{equation}
where \(\mathbf{z}_{\mathbf{i}}\in\mathbb{R}^{\mathbf{V}}\) represents the final single spatiotemporal joint representation vector output for subject \(\mathbf{i}\). This representation contains both the denoised features of spatial anatomy and the condensed longitudinal deterioration trajectory of the disease that varies across individuals.

\subsubsection{Longitudinal Graph Contrastive Supervision Based on InfoNCE}

In an unsupervised environment lacking clinical labels, this framework introduces a graph contrastive learning mechanism based on InfoNCE as the core engine driving the optimization and updating of the above networks.

Before computing the contrastive loss, the model introduces a nonlinear projection head (Projection Head, \(g(\cdot)\)) composed of a multilayer perceptron (MLP) to map the spatiotemporal representation into a lower-dimensional latent contrastive space:
\begin{equation}
\mathbf{p}_{\mathbf{i}}
=
g(\mathbf{z}_{\mathbf{i}})
=
\mathbf{W}_{\mathbf{2}}
\boldsymbol{\sigma}\!\left(
\mathbf{W}_{\mathbf{1}}
\mathbf{z}_{\mathbf{i}}
+
\mathbf{b}_{\mathbf{1}}
\right)
+
\mathbf{b}_{\mathbf{2}}
\tag{8}
\end{equation}
where \(\mathbf{W}_{\mathbf{1}}\in\mathbb{R}^{\mathbf{d}_{\mathbf{h}}\times \mathbf{V}}\) and \(\mathbf{W}_{\mathbf{2}}\in\mathbb{R}^{\mathbf{d}_{\mathbf{p}}\times \mathbf{d}_{\mathbf{h}}}\) are the weight parameter matrices of the projection head, respectively, \(\mathbf{d}_{\mathbf{h}}\) is the neuron dimension of the intermediate hidden layer of the projection head, \(\mathbf{d}_{\mathbf{p}}\) is the low-dimensional output dimension of the final latent contrastive space, \(\mathbf{b}_{\mathbf{1}}\) and \(\mathbf{b}_{\mathbf{2}}\) are bias vectors, and \(\mathbf{p}_{\mathbf{i}}\in\mathbb{R}^{\mathbf{d}_{\mathbf{p}}}\) is the coordinate vector of subject \(\mathbf{i}\) in the low-dimensional projection space. The low-dimensional mapping forces the network to strip redundant measurement noise and retain the core features representing individual disease identity. At the same time, it avoids the strong pull-push forces of contrastive learning directly destroying the true anatomical physical semantics contained in \(\mathbf{z}_{\mathbf{i}}\).

To construct contrastive positive and negative sample pairs without disrupting longitudinal temporal continuity, this model adopts an in-network feature perturbation mechanism. In a single training batch, the temporal imaging sequences of the same batch of subjects (batch size \(\mathbf{B}\)) continuously perform two forward passes in training mode. Since the graph convolutional encoder embeds a random dropout layer (Dropout, with a discard rate set to \(\mathbf{p}=\mathbf{0.5}\)), the two forward passes will adaptively activate different topological feature channels, thereby generating two independent views carrying weak spatial topological perturbations for subject \(\mathbf{i}\):

\textbf{View 1 (as Anchor):} outputs the spatiotemporal joint representation \(\mathbf{z}_{\mathbf{i}}^{(\mathbf{1})}\), which is mapped by the projection head to the latent space anchor coordinate \(\mathbf{p}_{\mathbf{i}}^{(\mathbf{1})}\in\mathbb{R}^{\mathbf{d}_{\mathbf{p}}}\), and simultaneously outputs the node-level reconstruction features \(\hat{\mathbf{X}}_{\mathbf{i},\mathbf{t}}^{(\mathbf{1})}\) at each time step;

\textbf{View 2 (as Positive Sample):} outputs the perturbed representation \(\mathbf{z}_{\mathbf{i}}^{(\mathbf{2})}\) of subject \(\mathbf{i}\) under a different dropout mask, corresponding to the latent space positive sample coordinate \(\mathbf{p}_{\mathbf{i}}^{(\mathbf{2})}\in\mathbb{R}^{\mathbf{d}_{\mathbf{p}}}\), and outputs the reconstruction features \(\hat{\mathbf{X}}_{\mathbf{i},\mathbf{t}}^{(\mathbf{2})}\);

\textbf{Negative Samples:} in the same batch, representations \(\mathbf{p}_{\mathbf{j}}^{(\mathbf{2})}\) (\(\mathbf{j}\neq \mathbf{i}\)) extracted from other different subjects in View 2 are strictly defined as negative samples.

In a training batch containing \(\mathbf{B}\) independent subjects, the temporal graph contrastive loss function \(\mathcal{L}_{\mathrm{InfoNCE}}\) is defined as follows:
\begin{equation}
\mathcal{L}_{\mathrm{InfoNCE}}
=
-\frac{\mathbf{1}}{\mathbf{B}}
\sum_{\mathbf{i}=\mathbf{1}}^{\mathbf{B}}
\log
\frac{
\exp\!\left(
\mathrm{sim}\!\left(
\mathbf{p}_{\mathbf{i}}^{(\mathbf{1})},
\mathbf{p}_{\mathbf{i}}^{(\mathbf{2})}
\right)/\boldsymbol{\tau}
\right)
}{
\exp\!\left(
\mathrm{sim}\!\left(
\mathbf{p}_{\mathbf{i}}^{(\mathbf{1})},
\mathbf{p}_{\mathbf{i}}^{(\mathbf{2})}
\right)/\boldsymbol{\tau}
\right)
+
\sum_{\mathbf{j}\neq \mathbf{i}}
\exp\!\left(
\mathrm{sim}\!\left(
\mathbf{p}_{\mathbf{i}}^{(\mathbf{1})},
\mathbf{p}_{\mathbf{j}}^{(\mathbf{2})}
\right)/\boldsymbol{\tau}
\right)
}
\tag{9}
\end{equation}

where \(\mathcal{L}_{\mathrm{InfoNCE}}\) is the overall graph contrastive learning loss value of the current batch, \(\mathbf{B}\) is the batch size during training, i.e., the number of subjects in a single optimization, \(\mathbf{i}\) and \(\mathbf{j}\) are traversal indices of subjects within the batch, where \(\mathbf{j}\neq \mathbf{i}\) indicates extraction of the corresponding negative sample, \(\mathbf{p}_{\mathbf{i}}^{(\mathbf{1})}\) is the anchor representation vector of subject \(\mathbf{i}\), \(\mathbf{p}_{\mathbf{i}}^{(\mathbf{2})}\) is the positive sample (perturbed) representation vector corresponding to subject \(\mathbf{i}\), \(\mathbf{p}_{\mathbf{j}}^{(\mathbf{2})}\) is the negative sample representation vector extracted from another different subject \(\mathbf{j}\), \(\mathrm{sim}(\mathbf{u},\mathbf{v})\) is the cosine similarity function between vectors, defined as \(\mathrm{sim}(\mathbf{u},\mathbf{v})=\frac{\mathbf{u}^{\top}\mathbf{v}}{\|\mathbf{u}\|_{\mathbf{2}}\|\mathbf{v}\|_{\mathbf{2}}}\), used to measure the angular distance between two representation vectors in the latent space, and \(\boldsymbol{\tau}\) is the temperature scaling hyperparameter, used to control the network's sensitivity to penalizing hard negative samples (different subjects that are too close in the feature space).

Meanwhile, to constrain the latent space from mode collapse and force the graph convolutional encoder to retain the underlying true anatomical measurement baseline, the model introduces a physical reconstruction error covering all \(\mathbf{T}\) time steps:
\begin{equation}
\mathcal{L}_{\mathrm{recon}}
=
\frac{\mathbf{1}}{\mathbf{B}\cdot \mathbf{T}}
\sum_{\mathbf{i}=\mathbf{1}}^{\mathbf{B}}
\sum_{\mathbf{t}=\mathbf{1}}^{\mathbf{T}}
\left(
\left\|
\hat{\mathbf{X}}_{\mathbf{i},\mathbf{t}}^{(\mathbf{1})}
-
\mathbf{X}_{\mathbf{i},\mathbf{t}}
\right\|_{\mathbf{F}}^{\mathbf{2}}
+
\left\|
\hat{\mathbf{X}}_{\mathbf{i},\mathbf{t}}^{(\mathbf{2})}
-
\mathbf{X}_{\mathbf{i},\mathbf{t}}
\right\|_{\mathbf{F}}^{\mathbf{2}}
\right)
\tag{10}
\end{equation}
where \(\mathbf{X}_{\mathbf{i},\mathbf{t}}\in\mathbb{R}^{\mathbf{V}\times \mathbf{1}}\) is the original input brain region volume of subject \(\mathbf{i}\) at time point \(\mathbf{t}\), and \(\|\cdot\|_{\mathbf{F}}\) is the Frobenius norm.

\begin{figure*}[t]   
\centering
\includegraphics[width=1\linewidth]{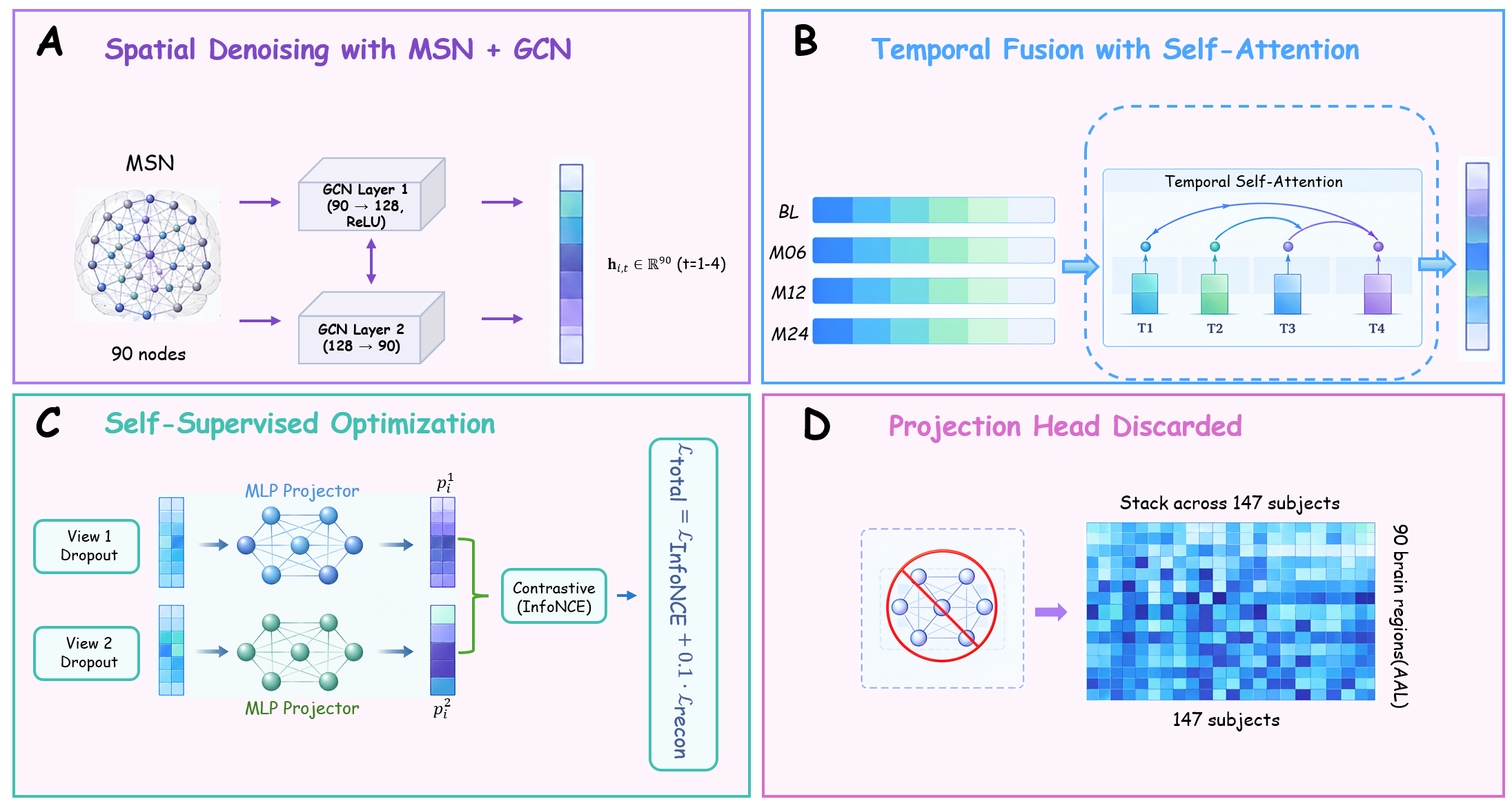}   
\caption{Detailed architecture of the spatiotemporal graph contrastive learning (ST-GCL) module. (A) Spatial denoising using morphological similarity network (MSN) and a two-layer graph convolutional network (GCN) to extract denoised regional features across four follow-up time points. (B) Temporal dynamic fusion aggregating longitudinal trajectories into a subject-level representation via temporal self-attention. (C) Self-supervised optimization via dropout-induced internal feature perturbation, guided by a joint loss $\mathcal{L}_{\text{total}} = \mathcal{L}_{\text{InfoNCE}} + 0.1 \cdot \mathcal{L}_{\text{recon}}$. (D) Feature extraction where the MLP projection head is discarded to retain anatomical interpretability, yielding the final spatiotemporal representation matrix $\mathbf{Y} \in \mathbb{R}^{147 \times 90}$. }
\label{fig_2}
\end{figure*}

The overall joint optimization objective of the front-end network is the weighted sum of the contrastive loss and the physical reconstruction error:
\begin{equation}
\mathcal{L}_{\mathrm{total}}
=
\mathcal{L}_{\mathrm{InfoNCE}}
+
\boldsymbol{\beta}
\mathcal{L}_{\mathrm{recon}}
\tag{11}
\end{equation}
where \(\boldsymbol{\beta}\) is a hyperparameter balancing the reconstruction constraint. After self-supervised training converges, the projection head \(g(\cdot)\) is discarded, and the purified spatiotemporal joint representation matrix \(\mathbf{Y}\in\mathbb{R}^{\mathbf{N}\times \mathbf{V}}\) output by the network, which has high longitudinal discriminative power, is fully exported and directly used as the imaging input for the downstream multi-omics association module.The framework of the front-end method is illustrated in Figure~\ref{fig_2}.

\subsection{Back End: Dual Graph-Regularized Multi-Task Sparse Canonical Correlation Analysis}
The front-end ST-GCL module successfully extracted highly discriminative spatiotemporal fused brain imaging representations from high-noise longitudinal MRI sequences. To further investigate the underlying microscopic molecular variations associated with this macroscopic atrophy trajectory, this study constructs a Dual Graph-Regularized Multi-Task Sparse Canonical Correlation Analysis (DG-MTSCCA) model at the back end. This module aims to precisely localize core pathogenic targets in genetic omics data through rigorous mathematical joint optimization, while simultaneously discovering their corresponding imaging susceptible brain regions \cite{ref24,ref25}.

\subsubsection{Mathematical Notation Definition and Graph Laplacian Topology Construction}
In this section, column vectors are denoted by bold lowercase letters, and matrices are denoted by bold uppercase letters. For a given matrix \(\mathbf{M}=(m_{\mathbf{ij}})\), \(\mathbf{m}^{\mathbf{i}}\) and \(\mathbf{m}_{\mathbf{j}}\) denote the \(\mathbf{i}\)-th row and \(\mathbf{j}\)-th column of the matrix, respectively. \(\|\mathbf{M}\|_{\mathbf{F}}=\sqrt{\sum_{\mathbf{i}}\sum_{\mathbf{j}}m_{\mathbf{ij}}^{\mathbf{2}}}\) denotes the Frobenius norm, \(\|\mathbf{M}\|_{\mathbf{2},\mathbf{1}}=\sum_{\mathbf{i}}\sqrt{\sum_{\mathbf{j}}m_{\mathbf{ij}}^{\mathbf{2}}}\) denotes the \(L_{\mathbf{2},\mathbf{1}}\) norm, and \(\|\mathbf{M}\|_{\mathbf{1}}=\sum_{\mathbf{i}}\sum_{\mathbf{j}}|m_{\mathbf{ij}}|\) denotes the \(L_{\mathbf{1}}\) norm.

This study treats different genetic omics as subtasks under a multi-task learning framework, and sets \(\mathbf{C}\) as the total number of genetic modalities (in this study \(\mathbf{C}=\mathbf{2}\), corresponding to SNP and DNA methylation, respectively).

\(\mathbf{X}_{\mathbf{c}}\in\mathbb{R}^{\mathbf{n}\times \mathbf{p}_{\mathbf{c}}}\): represents the feature matrix of the \(\mathbf{c}\)-th genetic modality (\(\mathbf{n}\) is the number of subjects, and \(\mathbf{p}_{\mathbf{c}}\) is the total number of variant loci in that modality).

\(\mathbf{Z}\in\mathbb{R}^{\mathbf{n}\times \mathbf{q}}\): represents the purified brain imaging spatiotemporal representation matrix output by the front-end ST-GCL.

\(\mathbf{u}_{\mathbf{c}}\in\mathbb{R}^{\mathbf{p}_{\mathbf{c}}}\): represents the canonical projection weight vector corresponding to the \(\mathbf{c}\)-th genetic modality. Let \(\mathbf{U}=[\mathbf{u}_{\mathbf{1}},\dots,\mathbf{u}_{\mathbf{C}}]\).

\(\mathbf{v}_{\mathbf{c}}\in\mathbb{R}^{\mathbf{q}}\): represents the imaging projection weight associated with the \(\mathbf{c}\)-th genetic modality. Let the global projection matrix of imaging features be \(\mathbf{V}=[\mathbf{v}_{\mathbf{1}},\dots,\mathbf{v}_{\mathbf{C}}]\in\mathbb{R}^{\mathbf{q}\times \mathbf{C}}\).

In molecular genetics, gene variants do not act in isolation. Significant linkage disequilibrium exists among SNPs, and methylation loci also form co-regulatory networks. To preserve this intrinsic biological topology during dimensionality reduction, this study constructs an undirected graph adjacency matrix \(\mathbf{W}_{\mathbf{c}}\) separately for each genetic modality (based on correlation coefficients among features with threshold filtering applied). Subsequently, the corresponding symmetric normalized graph Laplacian matrix \(\mathbf{L}_{\mathbf{c}}\) is derived:
\begin{equation}
\mathbf{L}_{\mathbf{c}}
=
\mathbf{I}
-
\mathbf{D}_{\mathbf{c}}^{-\mathbf{1}/\mathbf{2}}
\mathbf{W}_{\mathbf{c}}
\mathbf{D}_{\mathbf{c}}^{-\mathbf{1}/\mathbf{2}}
\tag{12}
\end{equation}

where \(\mathbf{I}\) is the identity matrix and \(\mathbf{D}_{\mathbf{c}}\) is the diagonal degree matrix (whose diagonal elements are the row sums of \(\mathbf{W}_{\mathbf{c}}\)). This Laplacian matrix \(\mathbf{L}_{\mathbf{c}}\) will be injected into the optimization model as a network topology penalty term.

\subsection{Objective Function Formulation}
The basic goal of DG-MTSCCA is to maximize the dual multivariate association between genetic data and multimodal imaging features in a low-dimensional latent space. By minimizing the reconstruction error in the projection space, its original optimization problem is defined as follows:
\begin{equation}
\begin{gathered}
\min_{\mathbf{u}_{\mathbf{c}},\mathbf{v}_{\mathbf{c}}}
\sum_{\mathbf{c}=\mathbf{1}}^{\mathbf{C}}
-
\mathbf{u}_{\mathbf{c}}^{\top}
\mathbf{X}_{\mathbf{c}}^{\top}
\mathbf{Z}
\mathbf{v}_{\mathbf{c}}
+
\Omega(\mathbf{U})
+
\Omega(\mathbf{V})
+
\boldsymbol{\gamma}
\sum_{\mathbf{c}=\mathbf{1}}^{\mathbf{C}}
\mathbf{u}_{\mathbf{c}}^{\top}
\mathbf{L}_{\mathbf{c}}
\mathbf{u}_{\mathbf{c}} \\[6pt]
\text{s.t.}\quad
\|\mathbf{X}_{\mathbf{c}}\mathbf{u}_{\mathbf{c}}\|_{\mathbf{2}}^{\mathbf{2}}
=
\|\mathbf{Z}\mathbf{v}_{\mathbf{c}}\|_{\mathbf{2}}^{\mathbf{2}}
=
\mathbf{1},\quad \forall \mathbf{c}
\end{gathered}
\tag{13}
\end{equation}

Under strict equality constraints, maximizing the inner product is equivalent to minimizing the Euclidean distance between the two. Therefore, Eq.~(13) can be rewritten in the following minimization form, with the sparsity penalty terms specified:
\begin{equation}
\begin{gathered}
\min_{\mathbf{u}_{\mathbf{c}},\mathbf{v}_{\mathbf{c}}}
\sum_{\mathbf{c}=\mathbf{1}}^{\mathbf{C}}
\|\mathbf{X}_{\mathbf{c}}\mathbf{u}_{\mathbf{c}}-\mathbf{Z}\mathbf{v}_{\mathbf{c}}\|_{\mathbf{2}}^{\mathbf{2}}
+
\boldsymbol{\lambda}_{\mathbf{1}}
\sum_{\mathbf{c}=\mathbf{1}}^{\mathbf{C}}
\|\mathbf{u}_{\mathbf{c}}\|_{\mathbf{1}}
+
\boldsymbol{\lambda}_{\mathbf{2}}
\|\mathbf{V}\|_{\mathbf{2},\mathbf{1}}
+
\boldsymbol{\gamma}
\sum_{\mathbf{c}=\mathbf{1}}^{\mathbf{C}}
\mathbf{u}_{\mathbf{c}}^{\top}
\mathbf{L}_{\mathbf{c}}
\mathbf{u}_{\mathbf{c}} \\[6pt]
\text{s.t.}\quad
\|\mathbf{X}_{\mathbf{c}}\mathbf{u}_{\mathbf{c}}\|_{\mathbf{2}}^{\mathbf{2}}
=
\|\mathbf{Z}\mathbf{v}_{\mathbf{c}}\|_{\mathbf{2}}^{\mathbf{2}}
=
\mathbf{1},\quad \forall \mathbf{c}
\end{gathered}
\tag{14}
\end{equation}

\begin{figure*}[t]
\centering
\includegraphics[width=1\linewidth]{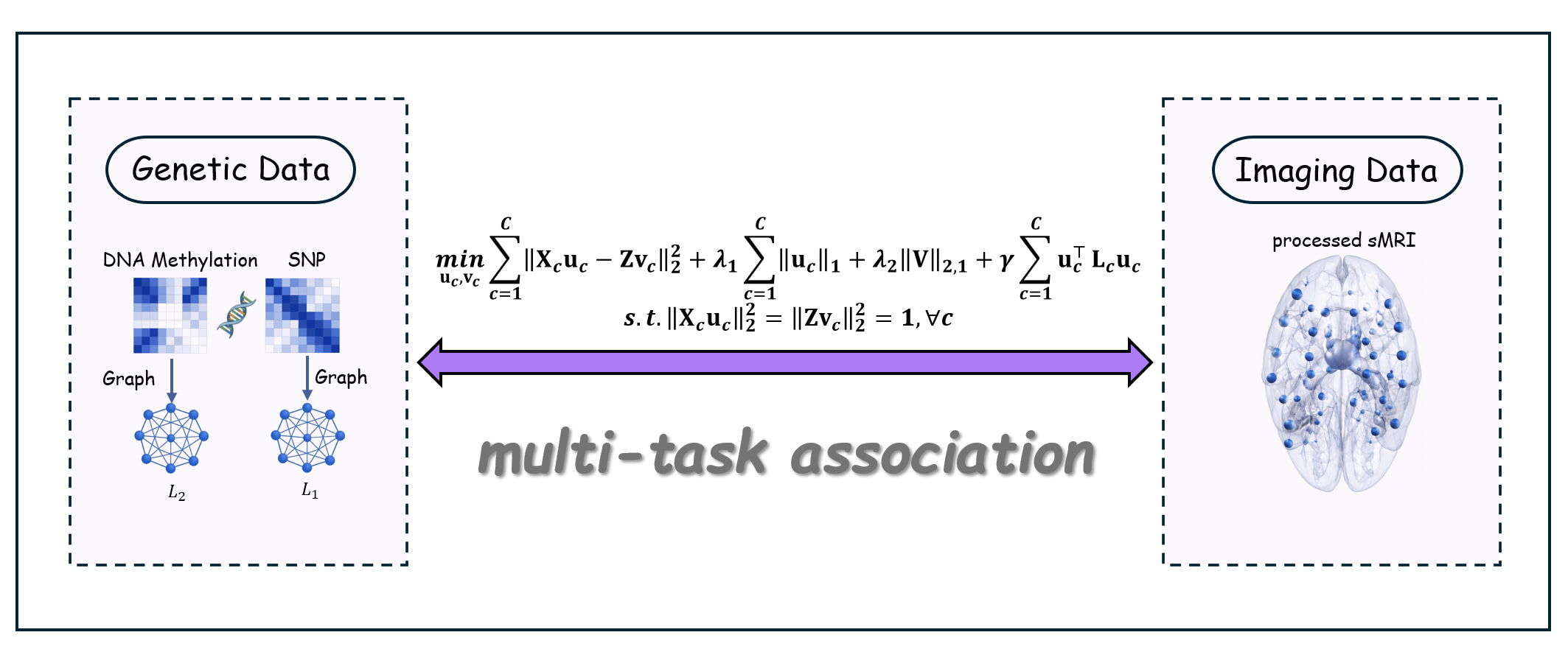}
\caption{Architecture of the dual graph-regularized multi-task sparse canonical correlation analysis (DG-MTSCCA) module.}
\label{fig_3}
\end{figure*}

Each constraint term in Eq.~(14) carries specific biological prior knowledge:

where the genetic specificity constraint \(\boldsymbol{\lambda}_{\mathbf{1}}\sum_{\mathbf{c}=\mathbf{1}}^{\mathbf{C}}\|\mathbf{u}_{\mathbf{c}}\|_{\mathbf{1}}\) induces strict element-level sparsity (\(L_{\mathbf{1}}\)-norm) among the high-dimensional and vast genetic variant loci, removes noise loci with no diagnostic value, and drives the model to select key variant loci under a specific modality, the imaging consistency constraint \(\boldsymbol{\lambda}_{\mathbf{2}}\|\mathbf{V}\|_{\mathbf{2},\mathbf{1}}\) imposes a group sparsity (\(L_{\mathbf{2},\mathbf{1}}\)-norm) penalty on the imaging weight matrix, forcing the projection weights of brain regions unrelated to disease progression to approach zero simultaneously across all genetic tasks \(\mathbf{c}\), thereby precisely localizing consistent susceptible brain regions jointly regulated by multiple genetic modalities, and the graph network guidance constraint \(\boldsymbol{\gamma}\sum_{\mathbf{c}=\mathbf{1}}^{\mathbf{C}}\mathbf{u}_{\mathbf{c}}^{\top}\mathbf{L}_{\mathbf{c}}\mathbf{u}_{\mathbf{c}}\) uses the graph Laplacian quadratic form, where when two genetic loci are highly correlated in the topological network, this term forces them to be assigned extremely similar coefficient values in the weight vector \(\mathbf{u}_{\mathbf{c}}\), thereby ensuring the coherence and robustness of feature selection along molecular biological pathways.The framework of the front-end method is illustrated in Figure~\ref{fig_3}.

\subsubsection{The Optimization Algorithm}
When \(\mathbf{U}\) and \(\mathbf{V}\) are considered jointly, solving Eq.~(14) is computationally challenging due to the non-convex bilinear interaction and spherical equality constraints. However, the objective can be decoupled into alternating convex subproblems: it is convex with respect to \(\mathbf{u}_{\mathbf{c}}\) when \(\mathbf{V}\) is fixed, and convex with respect to \(\mathbf{v}_{\mathbf{c}}\) when \(\mathbf{U}\) is fixed. Following established multi-task SCCA frameworks \cite{ref26,ref27,ref19}, this study employs an alternating iterative optimization algorithm. To address the optimization intractability of the hard equality constraints and ensure numerical non-singularity during matrix inversion, we reformulate each subproblem via Lagrangian relaxation by incorporating quadratic penalty terms \(\boldsymbol{\beta}_{\mathbf{u}}\|\mathbf{X}_{\mathbf{c}}\mathbf{u}_{\mathbf{c}}\|_{\mathbf{2}}^{\mathbf{2}}\) and \(\boldsymbol{\beta}_{\mathbf{v}}\|\mathbf{Z}\mathbf{v}_{\mathbf{c}}\|_{\mathbf{2}}^{\mathbf{2}}\). A subsequent normalization projection step is applied after each closed-form update to strictly satisfy the scale constraints and prevent trivial zero solutions.

\noindent\textbf{(1) Solving \(\mathbf{u}_{\mathbf{c}}\)}

First, all \(\mathbf{v}_{\mathbf{c}}\) are fixed, so that the objective function Eq.~(14) for a specific modality \(\mathbf{c}\) becomes:
\begin{equation}
\min_{\mathbf{u}_{\mathbf{c}}}
\left\|
\mathbf{X}_{\mathbf{c}}\mathbf{u}_{\mathbf{c}}
-
\mathbf{Z}\mathbf{v}_{\mathbf{c}}
\right\|_{\mathbf{2}}^{\mathbf{2}}
+
\boldsymbol{\lambda}_{\mathbf{1}}
\|\mathbf{u}_{\mathbf{c}}\|_{\mathbf{1}}
+
\boldsymbol{\gamma}
\mathbf{u}_{\mathbf{c}}^{\top}
\mathbf{L}_{\mathbf{c}}
\mathbf{u}_{\mathbf{c}}
+
\boldsymbol{\beta}_{\mathbf{u}}
\|\mathbf{X}_{\mathbf{c}}\mathbf{u}_{\mathbf{c}}\|_{\mathbf{2}}^{\mathbf{2}}
\tag{15}
\end{equation}
The derivative of this objective function with respect to \(\mathbf{u}_{\mathbf{c}}\) is computed and set to zero:
\begin{equation}
-
\mathbf{X}_{\mathbf{c}}^{\top}
\mathbf{Z}
\mathbf{v}_{\mathbf{c}}
+
\boldsymbol{\lambda}_{\mathbf{1}}
\tilde{\mathbf{D}}_{\mathbf{c}}
\mathbf{u}_{\mathbf{c}}
+
\boldsymbol{\gamma}
\mathbf{L}_{\mathbf{c}}
\mathbf{u}_{\mathbf{c}}
+
(\mathbf{1}+\boldsymbol{\beta}_{\mathbf{u}})
\mathbf{X}_{\mathbf{c}}^{\top}
\mathbf{X}_{\mathbf{c}}
\mathbf{u}_{\mathbf{c}}
=
\mathbf{0}
\tag{16}
\end{equation}

Here, since the \(L_{\mathbf{1}}\) norm is non-differentiable at zero, we introduce a diagonal matrix \(\tilde{\mathbf{D}}_{\mathbf{c}}\) for smoothing, whose \(\mathbf{i}\)-th diagonal element is \(\frac{\mathbf{1}}{\mathbf{2}|\mathbf{u}_{\mathbf{ci}}|+\boldsymbol{\epsilon}}\) (\(\mathbf{1}\leq \mathbf{i}\leq \mathbf{p}_{\mathbf{c}}\)). By factoring out \(\mathbf{u}_{\mathbf{c}}\), we can obtain its closed-form solution:
\begin{equation}
\mathbf{u}_{\mathbf{c}}
=
\left(
(\mathbf{1}+\boldsymbol{\beta}_{\mathbf{u}})
\mathbf{X}_{\mathbf{c}}^{\top}
\mathbf{X}_{\mathbf{c}}
+
\boldsymbol{\lambda}_{\mathbf{1}}
\tilde{\mathbf{D}}_{\mathbf{c}}
+
\boldsymbol{\gamma}
\mathbf{L}_{\mathbf{c}}
\right)^{-\mathbf{1}}
\mathbf{X}_{\mathbf{c}}^{\top}
\mathbf{Z}
\mathbf{v}_{\mathbf{c}}
\tag{17}
\end{equation}

Since the latent variable matrix \(\tilde{\mathbf{D}}_{\mathbf{c}}\) depends on \(\mathbf{u}_{\mathbf{c}}\) itself, this function can be easily solved by an iterative algorithm in the outer loop.

\noindent\textbf{(2) Solving \(\mathbf{v}_{\mathbf{c}}\)}

Next, the obtained \(\mathbf{U}\) is fixed, and the objective function Eq.~(14) for the imaging side can be written as:
\begin{equation}
\min_{\mathbf{v}_{\mathbf{c}}}
\sum_{\mathbf{c}=\mathbf{1}}^{\mathbf{C}}
\left\|
\mathbf{X}_{\mathbf{c}}\mathbf{u}_{\mathbf{c}}
-
\mathbf{Z}\mathbf{v}_{\mathbf{c}}
\right\|_{\mathbf{2}}^{\mathbf{2}}
+
\boldsymbol{\lambda}_{\mathbf{2}}
\|\mathbf{V}\|_{\mathbf{2},\mathbf{1}}
+
\boldsymbol{\beta}_{\mathbf{v}}
\sum_{\mathbf{c}=\mathbf{1}}^{\mathbf{C}}
\|\mathbf{Z}\mathbf{v}_{\mathbf{c}}\|_{\mathbf{2}}^{\mathbf{2}}
\tag{18}
\end{equation}

Taking its derivative with respect to \(\mathbf{v}_{\mathbf{c}}\) and setting it to zero yields:
\begin{equation}
-
\mathbf{Z}^{\top}
\mathbf{X}_{\mathbf{c}}
\mathbf{u}_{\mathbf{c}}
+
\boldsymbol{\lambda}_{\mathbf{2}}
\tilde{\mathbf{D}}_{\mathbf{v}}
\mathbf{v}_{\mathbf{c}}
+
(\mathbf{1}+\boldsymbol{\beta}_{\mathbf{v}})
\mathbf{Z}^{\top}
\mathbf{Z}
\mathbf{v}_{\mathbf{c}}
=
\mathbf{0}
\tag{19}
\end{equation}

Similarly, \(\tilde{\mathbf{D}}_{\mathbf{v}}\) is the diagonal matrix corresponding to the group sparse \(L_{\mathbf{2},\mathbf{1}}\) norm, whose \(\mathbf{j}\)-th diagonal element is \(\frac{\mathbf{1}}{\mathbf{2}\|\mathbf{v}^{\mathbf{j}}\|_{\mathbf{2}}+\boldsymbol{\epsilon}}\) (\(\mathbf{1}\leq \mathbf{j}\leq \mathbf{q}\)). We can further obtain:
\begin{equation}
\mathbf{v}_{\mathbf{c}}
=
\left(
(\mathbf{1}+\boldsymbol{\beta}_{\mathbf{v}})
\mathbf{Z}^{\top}
\mathbf{Z}
+
\boldsymbol{\lambda}_{\mathbf{2}}
\tilde{\mathbf{D}}_{\mathbf{v}}
\right)^{-\mathbf{1}}
\mathbf{Z}^{\top}
\mathbf{X}_{\mathbf{c}}
\mathbf{u}_{\mathbf{c}}
\tag{20}
\end{equation}

In the actual optimization procedure, the above closed-form solutions for \(\mathbf{u}_{\mathbf{c}}\) and \(\mathbf{v}_{\mathbf{c}}\) are alternately updated until a preset maximum number of iterations or a tolerance condition (\(\boldsymbol{\Delta}<\mathbf{10}^{-\mathbf{4}}\)) is reached. After convergence, the extracted non-zero \(\mathbf{u}_{\mathbf{c}}\) and \(\mathbf{V}\) represent the locked cross-modal pathogenic targets and affected brain regions.
\begin{algorithm}[htbp]
\caption{The DG-MTSCCA algorithm}
\label{alg:dg-mtscca}
\begin{algorithmic}[1]
\REQUIRE
The genetic data matrices \(\mathbf{X}_{\mathbf{c}}\in\mathbb{R}^{\mathbf{n}\times \mathbf{p}_{\mathbf{c}}}\), the longitudinal imaging phenotype data matrix \(\mathbf{Z}\in\mathbb{R}^{\mathbf{n}\times \mathbf{q}}\), \(\mathbf{c}\in\{\mathbf{1},\dots,\mathbf{C}\}\), the graph Laplacian matrices \(\mathbf{L}_{\mathbf{c}}\), and the pre-tuned hyperparameters \(\boldsymbol{\lambda}_{\mathbf{1}},\boldsymbol{\lambda}_{\mathbf{2}},\boldsymbol{\gamma},\boldsymbol{\beta}_{\mathbf{u}}\) and \(\boldsymbol{\beta}_{\mathbf{v}}\).
\ENSURE
Canonical weights \(\mathbf{U}\) and \(\mathbf{V}\).
\STATE Initialize \(\mathbf{u}_{\mathbf{c}}\in\mathbb{R}^{\mathbf{p}_{\mathbf{c}}}\), \(\mathbf{v}_{\mathbf{c}}\in\mathbb{R}^{\mathbf{q}}\) for \(\mathbf{c}\in\{\mathbf{1},\dots,\mathbf{C}\}\);
\WHILE{the convergence criterion is not met}
    \STATE Update the diagonal matrices \(\tilde{\mathbf{D}}_{\mathbf{c}}\) \((\mathbf{c}=\mathbf{1},\dots,\mathbf{C})\);
    \STATE Solve \(\mathbf{u}_{\mathbf{c}}\) according to Equation~(17), and scale \(\mathbf{u}_{\mathbf{c}}\) so that \(\|\mathbf{X}_{\mathbf{c}}\mathbf{u}_{\mathbf{c}}\|_{\mathbf{2}}=1\);
    \STATE Update the diagonal matrix \(\tilde{\mathbf{D}}_{\mathbf{v}}\);
    \STATE Solve \(\mathbf{v}_{\mathbf{c}}\) alternatively according to Equation~(20), and scale \(\mathbf{v}_{\mathbf{c}}\) so that \(\|\mathbf{Z}\mathbf{v}_{\mathbf{c}}\|_{\mathbf{2}}=1\);
\ENDWHILE
\end{algorithmic}
\end{algorithm}

In summary, to systematically analyze the brain structural evolution and multi-omics association mechanisms during the course of Alzheimer's disease, this study constructs a two-stage analysis system linking front-end representation purification with back-end association mining. In the front-end imaging analysis stage, relying on the spatiotemporal graph contrastive learning module and combining a brain region morphological similarity network with a dynamic temporal self-attention mechanism, spatial denoising and temporal condensation are performed on multi-time-point longitudinal MRI data of subjects, effectively extracting individual longitudinal evolution features with a high signal-to-noise ratio. In the back-end multimodal association stage, combining the dual graph-regularized multi-task sparse canonical correlation analysis module, a graph Laplacian matrix is constructed and introduced based on the internal correlation relationships among genetic features. Without relying on external biological databases, data-driven regularization constraints are used to preserve the intrinsic linkage disequilibrium and co-regulatory network structures within single nucleotide polymorphisms and DNA methylation, thereby mining the cross-omics covariance relationships between the feedforward imaging representations and high-dimensional genetic features. This study defines the overall technical route that links longitudinal imaging representations with multi-omics graph-regularized mining as the Spatiotemporal Graph Sparse Canonical Correlation Analysis framework (STGSCCA), providing a systematic solution for multimodal neuroimaging genetics research that accommodates both temporal evolution patterns and the intrinsic topological relationships of high-dimensional features.
\section{Data Source and Preprocessing}
\subsection{Clinical Cohort and Brain Imaging Data}
The clinical phenotypes, longitudinal structural magnetic resonance imaging (sMRI) data, and underlying genomic data used in this study were all obtained from the Alzheimer's Disease Neuroimaging Initiative (ADNI) database. To meet the requirements of longitudinal multi-omics joint analysis, this study performed cross-database matching and identified 147 subjects (comprising 91 EMCI and 56 CN subjects). Each subject has complete longitudinal sMRI scans across four follow-up time points (baseline, 6, 12, and 24 months), paired with baseline genome-wide SNP genotyping data and peripheral blood DNA methylation profiles \cite{ref21}.

For the longitudinal brain imaging data, this study used the SPM12 software pipeline for standardized spatial preprocessing, including image format conversion, gray matter/white matter segmentation based on tissue prior probability maps, bias field correction, and nonlinear registration to the standard Montreal Neurological Institute (MNI) space. Subsequently, the automated anatomical labeling (AAL) atlas was used to parcellate the brain into functional regions. After removing 32 cerebellar regions mainly responsible for low-level movement, gray matter volumes were finally extracted from 90 cortical and subcortical regions of interest (ROIs). The 90-dimensional feature sequences at these four time points were directly input into the front-end ST-GCL module for extracting spatiotemporal fusion trajectories \cite{ref27}. 

\subsection{Quality Control of Genetic Data}
For the microscopic genetic sequences, this study performed strict quality control (QC) before feature extraction:

\textbf{SNP data:} The PLINK toolkit was used for quality control of genome-wide association study (GWAS) data. Low-quality loci were removed, including those with minor allele frequency (MAF) \(<0.05\), those not conforming to Hardy-Weinberg equilibrium (HWE \(p<1.0\times 10^{-6}\)), and those with a sample missing rate higher than \(5\%\).

\textbf{DNA methylation data:} Methylation levels were quantified by \(\boldsymbol{\beta}\) values. The ChAMP package in R was used for signal normalization and batch effect correction of the raw data. For multiple overlapping \(\boldsymbol{\beta}\) value sequencing probes generated at the same CpG site for the same subject, this study calculated their average to enhance the numerical stability and consistency of the data \cite{ref5}.

After basic quality control, the original SNP and DNA methylation feature dimensions still reached the order of hundreds of thousands. Although SCCA theoretically supports solving in high-dimensional spaces where the number of features far exceeds the sample size, when omics features reach the order of \(10^{5}\) while the true clinical sample size is only 147, direct joint optimization faces enormous risks. First, the vast number of invalid features can easily induce the model to fit spurious correlations amid massive noise. Second, if tens of thousands of dimensions of omics data are directly aligned with only 90 dimensions of front-end brain imaging phenotypes, their enormous variance difference will completely dominate gradient updates, leading to severe modality imbalance. To address this, this study designed a strict pre-screening strategy, precisely truncating both types of omics features to 200 dimensions.

The genetic architecture underlying the Alzheimer's disease continuum involves both canonical risk loci with strong effect sizes and widely distributed polygenic variations. Consequently, for the SNP modality, we implemented a hybrid strategy comprising 80 prior-anchored features alongside 120 data-driven features. First, 80 candidate loci were anchored based on prior GWAS literature, providing targeted coverage of the established \textit{APOE} cluster and its high linkage disequilibrium (LD) region on chromosome 19 (including \textit{TOMM40} and \textit{NECTIN2}). These prior loci serve as biological positive controls (sanity checks), ensuring that the core etiologic baseline is retained against over-penalization. Subsequently, the remaining 120 slots were determined via data-driven screening. To strictly prevent data leakage across cross-validation folds, feature pre-screening was conducted solely within the training partition of each fold: genome-wide variants were ranked and selected according to their cross-covariance with the 90-dimensional baseline neuroimaging phenotypes.

For the high-dimensional DNA methylation data (Illumina HumanMethylation450), directly incorporating genome-wide probes risks overfitting and modality scale imbalance. Following standardized epigenomic quality control (removing cross-reactive probes, polymorphic CpGs, and probes with detection \(p > 0.01\)), an unsupervised feature pre-filtering strategy was adopted. Specifically, the top 200 CpG sites (\(P_{2} = 200\)) exhibiting the highest probe variability (standard deviation across samples) were retained. Importantly, this filtering was conducted in an entirely label-free, unsupervised manner without utilizing clinical diagnostic categories (EMCI vs.\ CN), thereby preventing label leakage into downstream classification tasks. Furthermore, fixing both SNP and DNA methylation dimensions to \(P_{1} = P_{2} = 200\) establishes structural symmetry in the feature space, avoiding optimization dominance caused by scale discrepancies during multi-task SCCA joint learning.
\begin{table}[htbp]
\centering
\caption{Clinical Information of Patients}
\label{tab:clinical}
\begin{tabularx}{\linewidth}{l C C C C}
\toprule
Groups & Number & Gender (M) & Gender (F) & Mean age \\
\midrule
EMCI & 91 & 50 & 41 & $70.31 \pm 5.88$ \\
CN   & 56 & 31 & 25 & $73.87 \pm 6.01$ \\
Total & 147 & 81 & 66 & $71.67 \pm 5.95$ \\
\bottomrule
\end{tabularx}
\end{table}

\subsection{Generation of Simulated Data}
To systematically evaluate the performance of the STGSCCA algorithm under a parameter-controlled environment, this study designed a data simulation framework based on a latent variable model. First, a unified ``ground truth'' was set, retaining only the first 15 features in each modality as true pathogenic targets, with true weights $\mathbf{w}^*$ linearly decreasing from $1.0$ to $0.3$ and the remaining background features set to zero. In the data generation stage, the subject cohort ($i = 1, \dots, n$) was driven by a shared individual-level latent variable $\mathbf{z} = [z_1, \dots, z_n]^\top \sim \mathcal{N}(\mathbf{0}, \mathbf{I})$. Baseline omics observation matrices were generated via noisy projections of this latent factor with the ground-truth weight vectors ($\mathbf{z}\mathbf{w}^{*\top}$ plus independent Gaussian noise $\mathcal{N}(0, \sigma^2)$). For the longitudinal neuroimaging sequence across $T = 4$ discrete follow-up time points ($t \in \{1, 2, 3, 4\}$), to realistically simulate inter-subject heterogeneity and nonlinear progression rates of neurodegenerative atrophy, the imaging observation vector $\mathbf{y}_{i,t}$ for subject $i$ at timepoint $t$ was generated by modulating the latent spatial signal as $\mathbf{y}_{i,t} = \phi_i(t) z_i \mathbf{w}_{\mathrm{MRI}}^* + \boldsymbol{\eta}_{i,t} + \boldsymbol{\epsilon}_{i,t}$. Here, $\phi_i(t) = \alpha_i \log(1 + t) + \beta_i t^2$ denotes the subject-specific nonlinear temporal decay function, where individual progression coefficients are independently sampled from $\alpha_i \sim \mathcal{N}(0.5, 0.05^2)$ and $\beta_i \sim \mathcal{N}(0.1, 0.02^2)$ to characterize individual variability in atrophy trajectories; $\boldsymbol{\eta}_{i,t} \sim \mathcal{N}(\mathbf{0}, \sigma_s^2 (\mathbf{I} + \mathbf{L}_0)^{-1})$ (with spatial perturbation scale $\sigma_s = 0.2$ and prior brain graph Laplacian $\mathbf{L}_0$) models spatially correlated graph perturbations across anatomically connected regions; and $\boldsymbol{\epsilon}_{i,t} \sim \mathcal{N}(\mathbf{0}, \sigma^2 \mathbf{I})$ represents scan-specific independent Gaussian measurement noise. Based on the above basic generation mechanism, this study constructed four groups of progressively escalating test sets by adjusting parameters. Data group 1, with $n = 300$ and $\sigma = 0.5$, served as an ideal baseline for establishing the upper limit of parameter recovery accuracy. Data group 2, with an extremely compressed sample size ($n = 80$ and $\sigma = 0.5$), was used to evaluate anti-overfitting capability in high-dimensional small-sample scenarios. Data group 3, with significantly amplified systematic noise ($n = 200$ and $\sigma = 2.5$), examined the signal capture and noise resistance robustness of the model under low signal-to-noise ratio conditions. Data group 4, maintaining standard cohort settings ($n = 300, \sigma = 0.5$), artificially injected a strong modular co-expression factor into the latent variable ($\mathbf{z}_{\mathrm{new}} = \mathbf{z} + 1.5\mathbf{z}_{\mathrm{module}}$, where $\mathbf{z}_{\mathrm{module}} \sim \mathcal{N}(\mathbf{0}, \mathbf{I})$) to verify the ability of the dual-graph Laplacian regularization term to identify and reconstruct highly collinear biomolecular pathways  \cite{ref29,ref30}.

 \section{Experimental Settings}
 \subsection{Experimental Parameter Settings}
In the front-end spatiotemporal fusion module, the model performs joint learning entirely in an unsupervised environment. To achieve effective feature disentanglement in the projection space, the contrastive learning InfoNCE loss and the physical reconstruction error are assigned a fixed weight ratio of 1.0 and 0.1, and the temperature hyperparameter \(\boldsymbol{\tau}\) is set to 0.1. The network is trained for 200 epochs using the Adam optimizer combined with a cosine annealing strategy. Finally, the longitudinal brain imaging sequences are reduced to a single 90-dimensional spatiotemporal representation matrix for use by the back end \cite{ref31}.

For the DG-MTSCCA, we use a nested 5-fold cross-validation method to find the optimal parameters. In the inner loop, the parameter combination that produces the highest average cross-task validation correlation coefficient is selected as the optimal hyperparameter, namely:
\begin{equation}
\mathrm{CV}(\boldsymbol{\lambda}_{\mathbf{1}},\boldsymbol{\lambda}_{\mathbf{2},\mathbf{1}},\boldsymbol{\gamma})
=
\frac{1}{5}
\sum_{\mathbf{j}=1}^{5}
\frac{1}{2}
\sum_{\mathbf{c}=1}^{2}
\mathrm{Corr}\!\left(
\bar{\mathbf{X}}_{\mathbf{c},\mathbf{j}}
\mathbf{u}_{\mathbf{c},\mathbf{j}},
\bar{\mathbf{Y}}_{\mathbf{j}}
\mathbf{v}_{\mathbf{c},\mathbf{j}}
\right)
\tag{21}
\end{equation}

where \(\bar{\mathbf{X}}_{\mathbf{c},\mathbf{j}}\) and \(\bar{\mathbf{Y}}_{\mathbf{j}}\) are the \(\mathbf{j}\)-th fold data used as the complementary test set for testing, and \(\mathbf{u}_{\mathbf{c},\mathbf{j}}\) and \(\mathbf{v}_{\mathbf{c},\mathbf{j}}\) are the canonical projection weights estimated from the training set.

To further reduce time overhead, we fixed the ridge regression penalty parameter \(\mathbf{\beta} = 0.5\). Since this parameter only plays a scaling role in the proximal gradient descent of the algorithm to ensure the numerical stability of matrix inversion, it only affects the magnitude of \(\mathbf{U}\) and \(\mathbf{V}\). This indicates that the relative sparse topological relationships among the variables remain unchanged. Finally, we tune the remaining core parameters, namely the \(\mathbf{L_1}\) penalty coefficient \(\mathbf{\lambda_1}\), the \(\mathbf{L_{2,1}}\) group sparsity coefficient \(\mathbf{\lambda_{2,1}}\), and the graph regularization coefficient \(\mathbf{\gamma}\), within the candidate range \(10^{\mathbf{i}}\) (\(\mathbf{i} = -4, -3, \dots, 0, \dots, 3, 4\)). Within this logarithmic interval, the model results fully cover the entire state space from less-sparsity to over-sparsity. Once the optimal parameters are determined in the inner loop, we refit the model on the outer training set and evaluate it on the held-out test set to obtain the final training and testing results.

In each alternating iteration experiment, all optimization methods stop if the projection weights of two adjacent updates satisfy the following condition:
\begin{equation}
\max_{\mathbf{i}}
\left|
\mathbf{u}_{\mathbf{c},\mathbf{i}}^{(t+1)}
-
\mathbf{u}_{\mathbf{c},\mathbf{i}}^{(t)}
\right|
\leq \boldsymbol{\epsilon}
\quad \text{and} \quad
\max_{\mathbf{j}}
\left|
\mathbf{v}_{\mathbf{c},\mathbf{j}}^{(t+1)}
-
\mathbf{v}_{\mathbf{c},\mathbf{j}}^{(t)}
\right|
\leq \boldsymbol{\epsilon}
\tag{22}
\end{equation}

where \(\boldsymbol{\epsilon}\) is the tolerance error. In the experiments of this paper, \(\boldsymbol{\epsilon}\) is empirically set to \(10^{-4}\). 
\subsection{Experimental Results on Simulated Data}
\begin{figure}[htbp]
\centering
\includegraphics[width=1.0\textwidth, height=0.95\textheight]{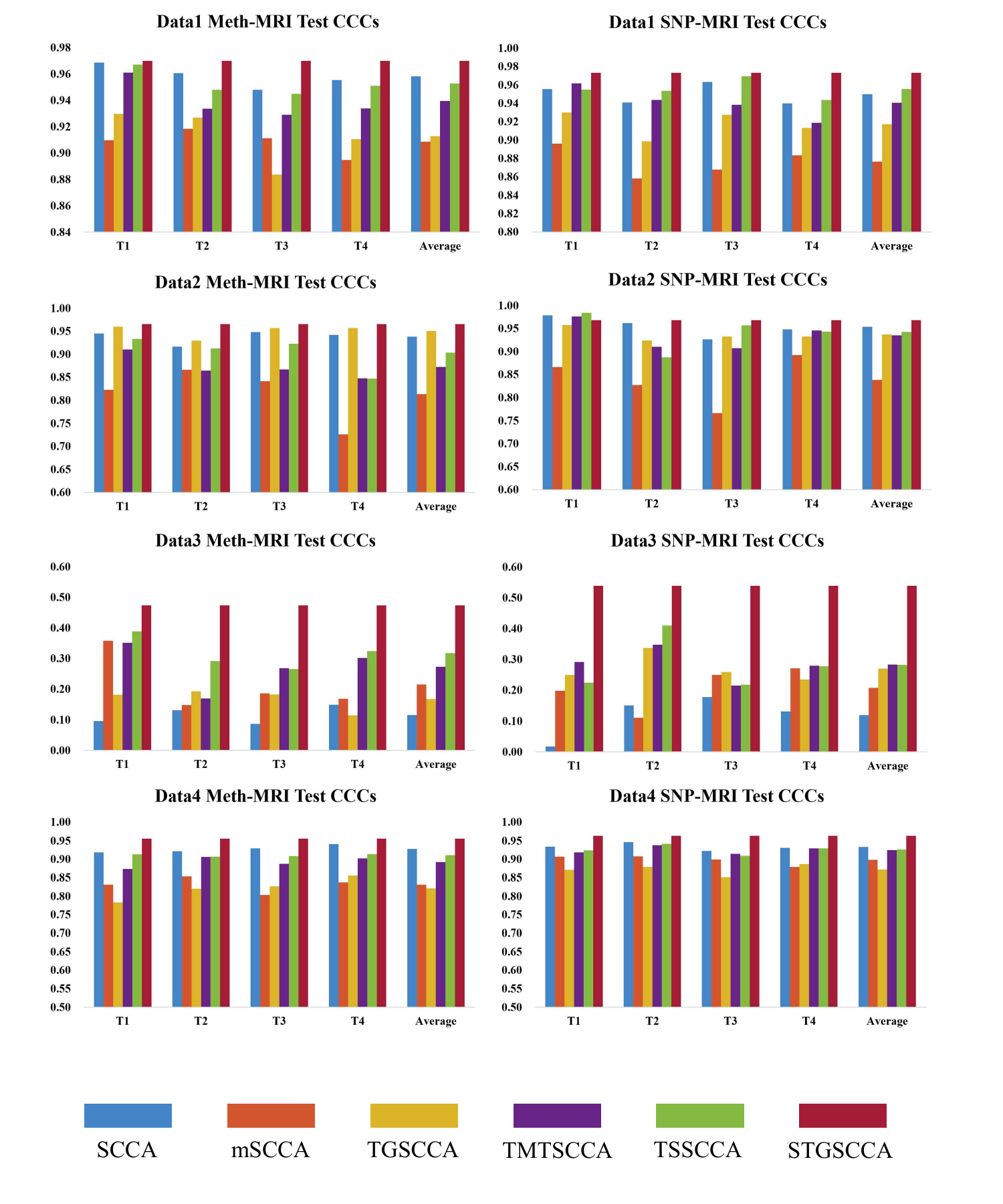}
\caption{CCC results on simulated data 1--4. The left side shows the association between imaging and DNA methylation, and the right side shows the association between imaging and SNP.}
\label{fig_4}
\end{figure}

To comprehensively evaluate the effectiveness of the proposed algorithm, this section systematically compares the STGSCCA framework with five algorithms, namely SCCA \cite{ref33}, mSCCA \cite{ref34}, TGSCCA \cite{ref18}, TMTSCCA \cite{ref19}, and TSSCCA \cite{ref21}, on four groups of simulated datasets with different challenging characteristics. These algorithms include base association algorithms and advanced longitudinal canonical correlation analysis algorithms, as shown in Figure~\ref{fig_4}. Limited by the length of the paper, and considering that the test set canonical correlation coefficient (Test CCC) can more truly and objectively reflect the generalization and anti-overfitting capability of the model on unseen data, this section uniformly omits the training set results. It is worth noting that in Figure~\ref{fig_4}, traditional longitudinal algorithms output fluctuating CCC values at four independent time points, from \(T_{1}\) to \(T_{4}\), whereas STGSCCA shows completely consistent scores at each time point and in the average value. This phenomenon is not computational redundancy. Rather, it is because the front-end ST-GCL module of this model has broken the isolated temporal barriers and perfectly compressed the four longitudinal follow-up sequences into a single spatiotemporal fused representation with a global view. Therefore, when this representation is aligned with omics data at the back end, what is extracted is an overall evolution pattern that encompasses the whole picture, thereby producing a single, unified canonical correlation coefficient (CCC) representing the overall longitudinal profile, which is presented across time points for visual comparison. From the comparison results across the four scenarios, STGSCCA maintains competitive and stable generalization performance. In the conventional sample scenario (Data 1) and the small-sample scenario (Data 2), compared with the performance fluctuations of baseline algorithms under high dimensionality, STGSCCA exhibits consistent anti-overfitting capability aided by the regularization effect of front-end contrastive learning. In the high-noise environment (Data 3), where competing methods experience substantial performance degradation, STGSCCA attenuates random noise via spatial smoothing and temporal self-attention, retaining higher signal recovery. Moreover, in the collinearity environment with modular co-expression factors (Data 4), the dual-graph Laplacian regularization term in STGSCCA mitigates the tendency of standard $L_1$ penalties to arbitrarily discard correlated targets, achieving a test association score above 0.95.

To evaluate the true parameter recovery capability under controlled conditions, we benchmarked STGSCCA against five classical methods across four distinct synthetic scenarios (Data 1--4). Rather than using binary classification metrics that rely on subjective zero-thresholding and frequently induce artificial saturation, we directly measure the geometric recovery error of estimated canonical weights against the predefined ground truth.

In each synthetic benchmark, the ground-truth canonical weight vector is denoted as $\mathbf{w}^{*} \in \mathbb{R}^{p}$. Specifically, true pathogenic signals are assigned to the first $k = 15$ biomarkers with linearly decaying effect sizes ($w_{j}^{*} = 1.0 - 0.05(j - 1)$ for $j = 1, \dots, 15$), while the remaining background features represent uninformative noise ($w_{j}^{*} = 0$ for $j > 15$). To quantify the estimation error, we compute the unit-normalized Weight Root Mean Square Error (Weight RMSE):
\begin{equation}
    \mathrm{Weight\ RMSE}(\hat{\mathbf{w}}, \mathbf{w}^{*}) = \left\| \frac{\hat{\mathbf{w}}}{\|\hat{\mathbf{w}}\|_{2}} - \frac{\mathbf{w}^{*}}{\|\mathbf{w}^{*}\|_{2}} \right\|_{2}
    \tag{23}
\end{equation}
where $\hat{\mathbf{w}} \in \mathbb{R}^{p}$ represents the estimated canonical weight vector aggregated across time points (i.e., $\hat{\mathbf{w}} = \frac{1}{T}\sum_{t=1}^{T} |\mathbf{w}_{t}|$ for longitudinal imaging), and $\|\cdot\|_{2}$ denotes the standard Euclidean norm. Normalizing by the $\ell_{2}$ norm eliminates the scale indeterminacy inherent to canonical correlation models, projecting both vectors onto a unit hypersphere for unbiased comparison. The resulting metric strictly ranges within $[0, \sqrt{2}]$, where $0$ corresponds to perfect parameter recovery and $\sqrt{2} \approx 1.414$ marks an entirely orthogonal (uninformative) parameter profile.

\begin{figure}[t]
\centering
\includegraphics[width=\textwidth]{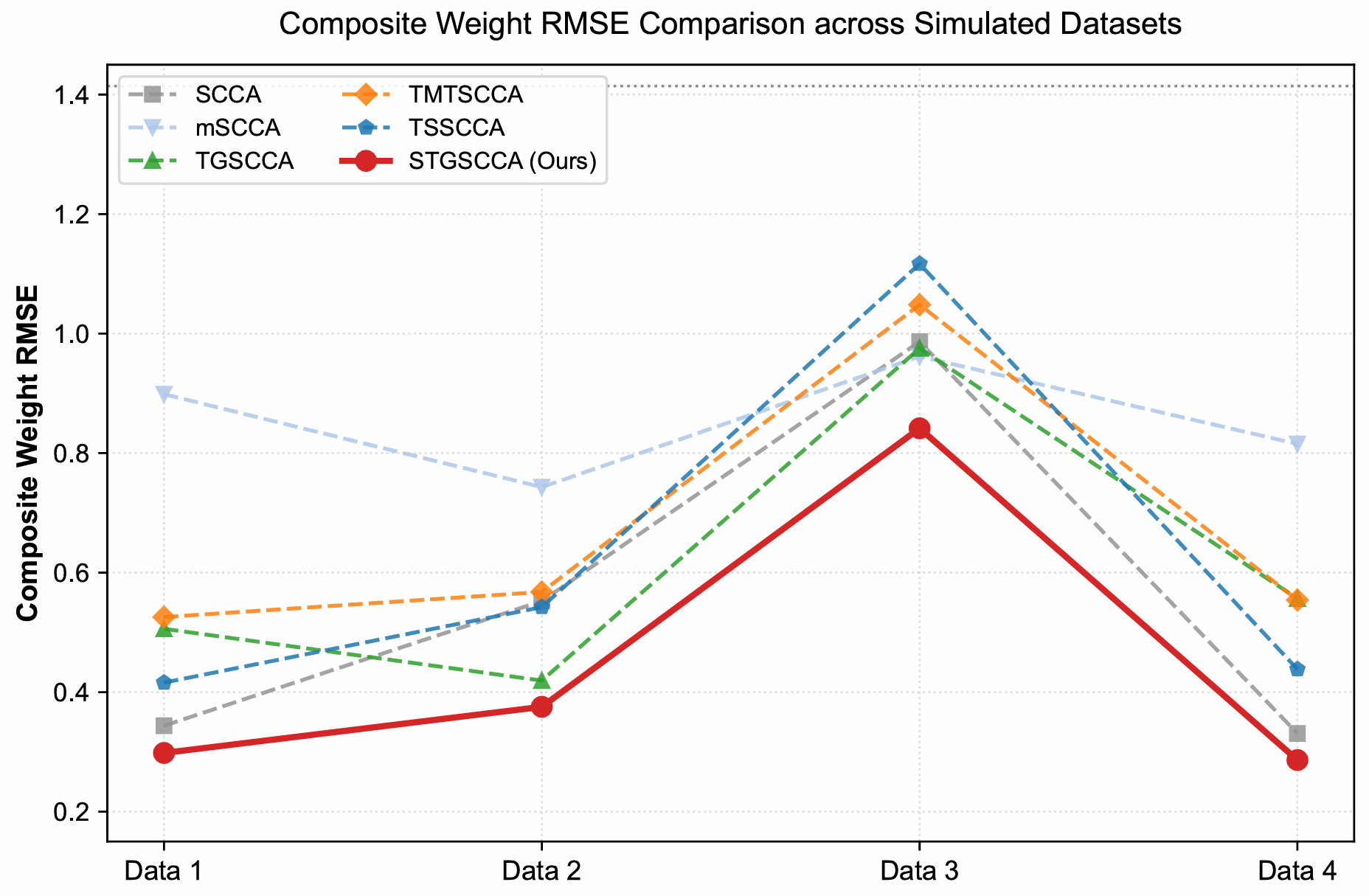}
\caption{Composite Weight RMSE comparison across four simulated datasets: Data 1 (Standard benchmark: $n=300, \sigma=0.5$), Data 2 (High-dimensional small-sample: $n=80, \sigma=0.5$), Data 3 (Extreme Gaussian noise: $n=200, \sigma=2.5$), and Data 4 (Modular graph topology: $n=300, \sigma=0.5$). The dotted horizontal line marks the theoretical uninformative orthogonal limit ($\sqrt{2} \approx 1.414$). Lower values denote superior parameter reconstruction fidelity.}
\label{fig_5}
\end{figure}

For multi-omics assessment, the \textbf{Composite Weight RMSE} is defined as the arithmetic mean of reconstruction errors across the three biological modalities:
\begin{equation}
    \mathrm{Composite\ Weight\ RMSE} = \frac{1}{3} \sum_{m \in \{\mathrm{DNAm}, \mathrm{SNP}, \mathrm{MRI}\}} \mathrm{Weight\ RMSE}\left(\hat{\mathbf{w}}^{(m)}, \mathbf{w}^{*(m)}\right).
    \tag{24}
\end{equation}
Figure~\ref{fig_5} depicts the evolutionary trajectories of the Composite Weight RMSE across the four synthetic benchmark environments. Overall, STGSCCA consistently maintains the lowest estimation error across all testing scenarios, while tracking the performance trajectories of competitive baselines realistically. In the standard baseline setting (Data 1), STGSCCA achieves a composite RMSE of $0.2982$, providing a competitive edge over regularized baselines such as SCCA ($0.3438$) and TSSCCA ($0.4159$).

Under small-sample stress (Data 2, $n=80$), all models experience expected variance inflation; however, STGSCCA curtails the performance penalty ($\mathrm{RMSE} = 0.3754$), benefiting from the dual-graph Laplacian priors that stabilize coefficient paths against overfitting. When subjected to severe measurement noise (Data 3, $\sigma=2.5$), competing models degrade toward the uninformative boundary ($\mathrm{RMSE} \approx 0.96\text{--}1.11$), whereas STGSCCA exhibits stronger noise attenuation ($\mathrm{RMSE} = 0.8415$). Finally, in the presence of modular collinearity (Data 4), STGSCCA reconstructs correlated functional blocks smoothly ($\mathrm{RMSE} = 0.2863$), outperforming sparse counterparts that randomly drop co-expressed features.

\subsection{Experimental Results on Real ADNI Data}
\subsubsection{Association Experiment Results}
To verify the generalization capability and application potential of the proposed framework in a complex real clinical environment, this section conducted cross-modal association analysis on the real Alzheimer's Disease Neuroimaging Initiative (ADNI) cohort, with results shown in Table~\ref{tab_2}. Unlike highly idealized simulated data, real imaging genetics data is accompanied by extremely high clinical heterogeneity and massive systematic background noise, causing the test set canonical correlation coefficients (Test CCCs) of all algorithms to show a reasonable decline compared with the simulation experiments. Despite facing severe data distribution challenges, STGSCCA shows competitive performance on the two core tasks. For Meth-MRI joint analysis, it exceeds the 0.22 performance level of state-of-the-art longitudinal algorithms including TMTSCCA and TSSCCA, achieving the highest average score of \(0.26\pm 0.01\). In the SNP-MRI association task, which has higher dimensionality and weaker genetic penetrance, it likewise ranks first with an average score of \(0.21\pm 0.05\). It is worth noting that, as described in the simulated data experiment section above, STGSCCA presents completely consistent association scores at the four time points \(T_{1}\) to \(T_{4}\). This is because the front-end ST-GCL module has already fused the longitudinal follow-up sequences into a single global spatiotemporal representation in advance, yielding a consolidated longitudinal association score (CCC) when aligned with omics data at the back end. To ensure consistency in the figure comparison, it is mapped in parallel to each time point for display. The experimental results on the real ADNI cohort strongly prove that the end-to-end architecture of STGSCCA, which performs spatiotemporal fusion first and cross-modal alignment second, can effectively avoid the clinical noise interference of a single follow-up, and, with the help of the dual graph regularization mechanism, robustly lock onto the most disease-specific association targets from the vast omics space.
\begin{table}[!htbp]
\centering
\caption{Cross-modal association performance (Test CCCs) of different methods on the real ADNI cohort across four follow-up time points.}
\label{tab_2}
\small
\setlength{\tabcolsep}{5.5pt}
\begin{tabular}{llccccc}
\toprule
\textbf{Task} & \textbf{Methods} & $T_1$ & $T_2$ & $T_3$ & $T_4$ & \textbf{Average} \\
\midrule
\multirow{6}{*}{\textbf{Meth--MRI}} 
 & SCCA     & $0.19\pm0.02$ & $0.21\pm0.00$ & $0.20\pm0.00$ & $0.15\pm0.01$ & $0.18\pm0.01$ \\
 & mSCCA    & $0.23\pm0.05$ & $0.23\pm0.01$ & $0.20\pm0.06$ & $0.21\pm0.00$ & $0.21\pm0.00$ \\
 & TGSCCA   & $0.17\pm0.01$ & $0.18\pm0.04$ & $0.17\pm0.00$ & $0.17\pm0.03$ & $0.17\pm0.01$ \\
 & TMTSCCA  & $0.21\pm0.07$ & $0.23\pm0.00$ & $0.21\pm0.01$ & $0.22\pm0.01$ & $0.22\pm0.03$ \\
 & TSSCCA   & $0.23\pm0.01$ & $0.24\pm0.00$ & $0.21\pm0.04$ & $0.21\pm0.01$ & $0.22\pm0.00$ \\
 & \textbf{STGSCCA} & $\mathbf{0.26\pm0.01}$ & $\mathbf{0.26\pm0.01}$ & $\mathbf{0.26\pm0.01}$ & $\mathbf{0.26\pm0.01}$ & $\mathbf{0.26\pm0.01}$ \\
\midrule
\multirow{6}{*}{\textbf{SNP--MRI}} 
 & SCCA     & $0.11\pm0.07$ & $0.11\pm0.03$ & $0.12\pm0.00$ & $0.13\pm0.06$ & $0.12\pm0.01$ \\
 & mSCCA    & $0.18\pm0.04$ & $0.13\pm0.00$ & $0.14\pm0.08$ & $0.15\pm0.05$ & $0.15\pm0.00$ \\
 & TGSCCA   & $0.10\pm0.03$ & $0.10\pm0.02$ & $0.13\pm0.07$ & $0.11\pm0.08$ & $0.11\pm0.05$ \\
 & TMTSCCA  & $0.20\pm0.08$ & $\mathbf{0.22\pm0.02}$ & $0.21\pm0.00$ & $0.20\pm0.05$ & $0.20\pm0.04$ \\
 & TSSCCA   & $0.20\pm0.01$ & $0.21\pm0.00$ & $\mathbf{0.22\pm0.08}$ & $0.19\pm0.00$ & $0.20\pm0.00$ \\
 & \textbf{STGSCCA} & $\mathbf{0.21\pm0.05}$ & $0.21\pm0.05$ & $0.21\pm0.05$ & $\mathbf{0.21\pm0.05}$ & $\mathbf{0.21\pm0.05}$ \\
\bottomrule
\end{tabular}
\vspace{2pt}
\begin{flushleft}
\footnotesize{Note: Values are expressed as mean $\pm$ standard deviation (SD) across 5-fold cross-validation. Bold values denote the highest numerical performance in each column.}
\end{flushleft}
\end{table}

\subsubsection{Feature Selection Results}
To evaluate the biological relevance and clinical interpretability of the features extracted by the proposed algorithm for early mild cognitive impairment (EMCI) profiling, this section conducts an in-depth analysis of the top 10 cross-modal candidate targets identified by each method on the real-world ADNI cohort, with comparative results summarized in Tables~\ref{tab_3}--\ref{tab_4}. For the longitudinal baseline algorithms, absolute canonical projection weights across the four consecutive follow-up time points were arithmetically averaged and ranked in descending order to select the top 10 markers. In contrast, STGSCCA leverages its front-end spatiotemporal representation module to derive a unified, subject-level weight vector that integrates longitudinal morphological trajectories, thereby capturing coordinated multi-omics signatures across the disease continuum. Functional annotation and literature comparisons demonstrate that the features prioritized by STGSCCA are predominantly enriched in biological pathways associated with neurodegenerative vulnerability along the Alzheimer's disease (AD) continuum, aligning with the subtle pathological alterations observed in EMCI. Specifically, in the SNP modality, the model reliably prioritizes high-weight variants located within the \textit{APOE} (rs405509) \cite{ref35,ref36} and \textit{TOMM40} (rs1160985, rs8106922, rs405697, rs157580) \cite{ref37} gene clusters. These loci constitute foundational genetic risk baselines that modulate individual susceptibility to cognitive decline in aging populations. In the DNA methylation modality, STGSCCA consistently maps to core epigenetic markers associated with early neurodegenerative pathology, including \textit{BIN1} (cg06021088) \cite{ref38}, \textit{APOE} (cg06750524) \cite{ref39}, \textit{TREM2} (cg25748868) \cite{ref40}, and \textit{CLU} (cg00359590) \cite{ref41}. Although competitive longitudinal models such as TMTSCCA and TSSCCA capture prominent isolated targets like \textit{APOE} (rs405509), their lack of graph regularization to preserve endogenous biological topologies limits their sensitivity to cooperative immune and metabolic regulators (e.g., \textit{TREM2} and \textit{CLU}), resulting in top-ranked candidate sets confounded by non-specific background variations.

Beyond canonical risk loci, the top 10 feature set of STGSCCA captures several exploratory candidate markers whose functions in early-stage cognitive impairment warrant further investigation, including \textit{PTPN7} (cg03649293) \cite{ref42}, \textit{COL9A3} (cg15505615) \cite{ref43}, \textit{BCAM} (rs28399659, rs28399658) \cite{ref44}, and \textit{CNTN4} (rs2728067) \cite{ref45}. These selections emerge from coordinated cross-modal covariance patterns identified across genome-wide variations, epigenetic states, and longitudinal brain structural changes. By coupling temporal attention pooling with dual-graph Laplacian regularization, STGSCCA isolates subtle synergistic signals that are typically obscured by unconstrained sparse models, highlighting candidate molecular components that may modulate neurodegenerative susceptibility during the prodromal phase. Identifying these coordinated multi-omics profiles provides data-driven hypotheses for unraveling the biological heterogeneity of EMCI and prioritizing potential candidate targets for early clinical evaluation.

\begin{figure}[htbp]
\centering
\includegraphics[width=\textwidth]{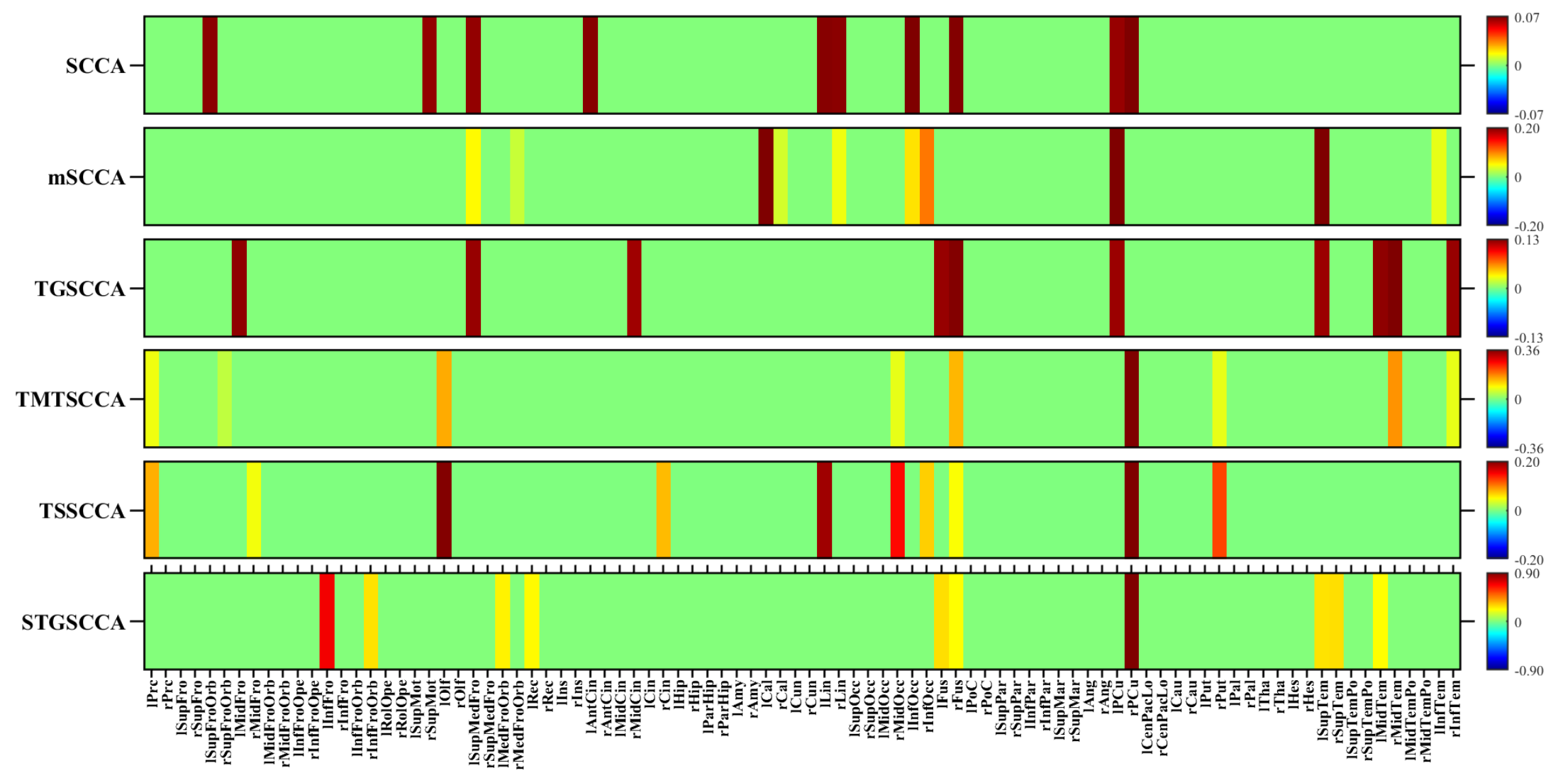}
\caption{Top 10 ROIs identified by longitudinal time methods. Each row corresponds to an SCCA method: (1) SCCA; (2) mSCCA; (3) TGSCCA; (4) TMTSCCA; (5) TSSCCA; (6) STGSCCA.}
\label{fig_6}
\end{figure}

\begin{figure}[htbp]
\centering
\includegraphics[width=\textwidth]{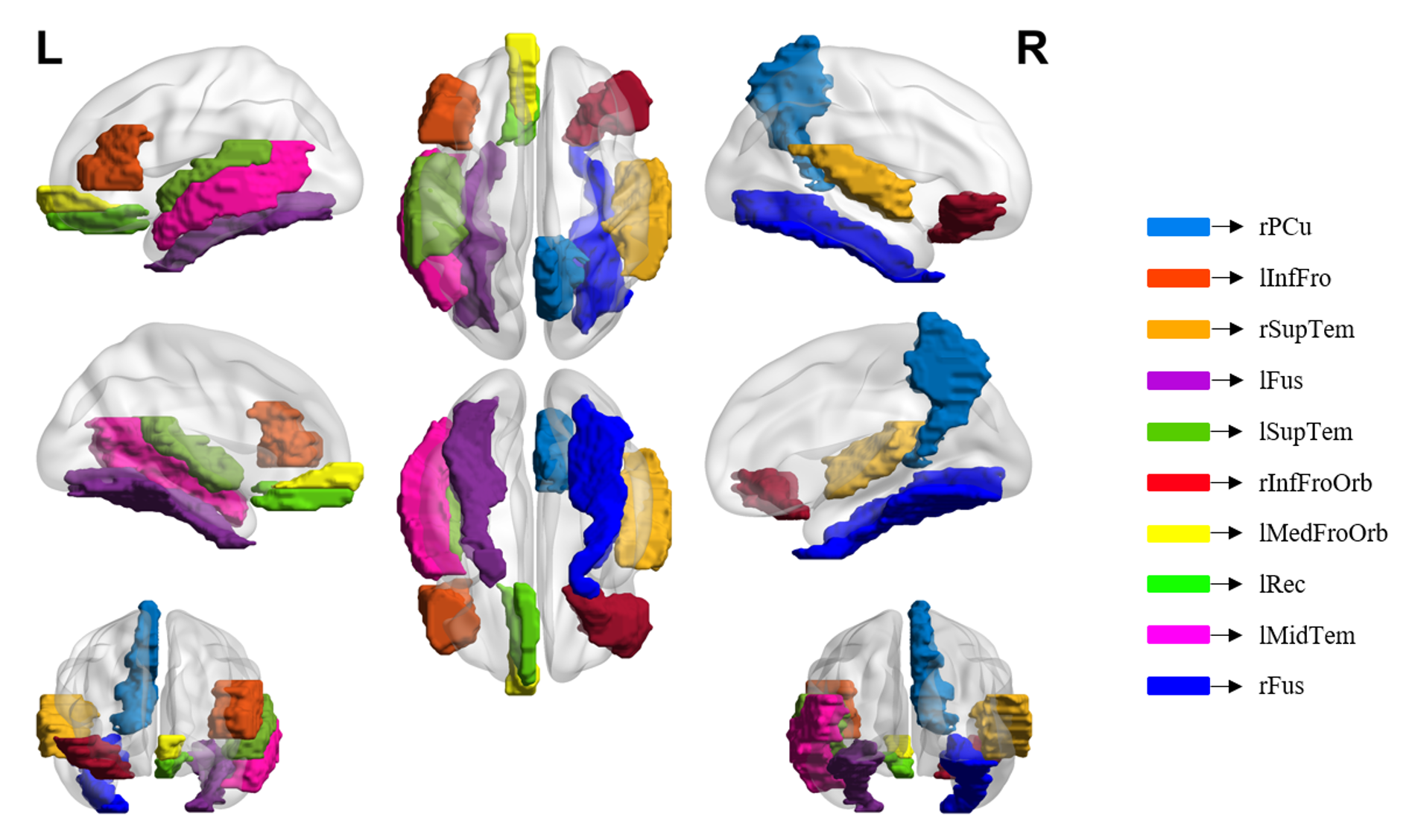}
\caption{Top 10 ROIs identified by STGSCCA.}
\label{fig_7}
\end{figure}

To assess the biological validity of the framework at the macroscopic anatomical scale, we evaluated the top 10 regions of interest (ROIs) prioritized across models on the ADNI cohort (Figure~\ref{fig_6}). In parallel with the omics analysis, baseline longitudinal weights were averaged across the four consecutive follow-up time points for descending ranking, whereas STGSCCA outputs a consolidated projection vector representing the cumulative spatiotemporal atrophy pattern. As illustrated in the 3D cortical mapping (Figure~\ref{fig_7}), STGSCCA localizes a coordinated pattern of regional vulnerability associated with cognitive impairment in EMCI. The top 10 prioritized brain regions comprise: the right precuneus (rPCu), left inferior frontal gyrus (lInfFro), bilateral superior temporal gyrus (rSupTem, lSupTem), bilateral fusiform gyrus (lFus, rFus), right inferior frontal gyrus, orbital part (rInfFroOrb), left medial superior frontal gyrus, orbital part (lMedFroOrb), left gyrus rectus (lRec), and left middle temporal gyrus (lMidTem). From a neuroanatomical perspective, the precuneus (rPCu) serves as a principal hub within the default mode network (DMN) \cite{ref46}, exhibiting early metabolic vulnerability and structural disconnectivity along the AD continuum, thereby providing critical morphological indications during the EMCI phase. Concurrently, regional volume alterations in the temporal structures and fusiform gyrus show documented associations with episodic memory decline and associative visual processing deficits observed in cognitively impaired cohorts. While comparative longitudinal methods capture signals in the right precuneus, their weight distributions remain dispersed and include primary sensory cortices, such as the inferior occipital gyrus and lingual gyrus, which exhibit low diagnostic specificity for early cognitive impairment. In contrast, STGSCCA reconstructs a coordinated temporo-parieto-frontal network characteristic of early structural decline in EMCI, demonstrating that spatiotemporal representation learning suppresses scan-specific acquisition noise and isolates neuroimaging markers consistent with modern neuropathological consensus.

\begin{table}[htbp]
\centering
\caption{Top 10 CpG sites identified by longitudinal methods}
\label{tab_3}
\begin{tabular}{lcccccc}
\toprule
Method & SCCA & mSCCA & TGSCCA & TMTSCCA & TSSCCA & STGSCCA \\
\midrule
NO.1  & cg09632220 & cg21576904 & cg09632220 & cg03552804 & cg10095954 & cg06021088 \\
NO.2  & cg23360102 & cg26859878 & cg10095954 & cg10095954 & cg15505615 & cg06750524 \\
NO.3  & cg26434983 & cg03884592 & cg03552804 & cg26859878 & cg16634253 & cg25748868 \\
NO.4  & cg21576904 & cg10095954 & cg20862251 & cg09632220 & cg21576904 & cg24472477 \\
NO.5  & cg08079330 & cg03649293 & cg10256162 & cg21576904 & cg06750524 & cg02856402 \\
NO.6  & cg06204447 & cg04400654 & cg26859878 & cg23762517 & cg05675373 & cg03649293 \\
NO.7  & cg26859878 & cg06204447 & cg23046711 & cg15911328 & cg23762517 & cg00359590 \\
NO.8  & cg04400654 & cg01382567 & cg10383396 & cg03649293 & cg24500823 & cg15505615 \\
NO.9  & cg11298786 & cg23046711 & cg25041922 & cg23360102 & cg08079330 & cg15185804 \\
NO.10 & cg04222159 & cg12151382 & cg15911328 & cg22366698 & cg26859878 & cg12151382 \\
\bottomrule
\end{tabular}
\end{table}

\begin{table}[htbp]
\centering
\caption{Top 10 SNPs identified by longitudinal methods}
\label{tab_4}
\begin{tabular}{lcccccc}
\toprule
Method & SCCA & mSCCA & TGSCCA & TMTSCCA & TSSCCA & STGSCCA \\
\midrule
NO.1  & rs11711623 & rs42451    & rs11129295 & rs11129295 & rs164215    & rs28399659 \\
NO.2  & rs2138506  & rs2965174  & rs42451    & rs405509   & rs42451     & rs28399658 \\
NO.3  & rs7638616  & rs2965101  & rs13088044 & rs164215   & rs11129295  & rs405509   \\
NO.4  & rs12485277 & rs1004165  & rs1947238  & rs986119   & rs7638616   & rs1160985  \\
NO.5  & rs6773381  & rs986119   & rs1160985  & rs6773381  & rs405509    & rs8106922  \\
NO.6  & rs2965174  & rs164215   & rs2122369  & rs2163842  & rs2163842   & rs6773439  \\
NO.7  & rs10416445 & rs387976   & rs4683071  & rs4803760  & rs4683071   & rs6769249  \\
NO.8  & rs7639787  & rs203713   & rs2965174  & rs387976   & rs2974384   & rs405697   \\
NO.9  & rs7644439  & rs3786507  & rs1020997  & rs13083647 & rs13083647  & rs157580   \\
NO.10 & rs6766086  & rs12610605 & rs164215   & rs7638616  & rs986119    & rs2728067  \\
\bottomrule
\end{tabular}
\end{table}

\subsubsection{Analysis of the Identified Joint Gene Set}
\begin{figure}[htbp]
\centering
\includegraphics[width=0.85\textwidth]{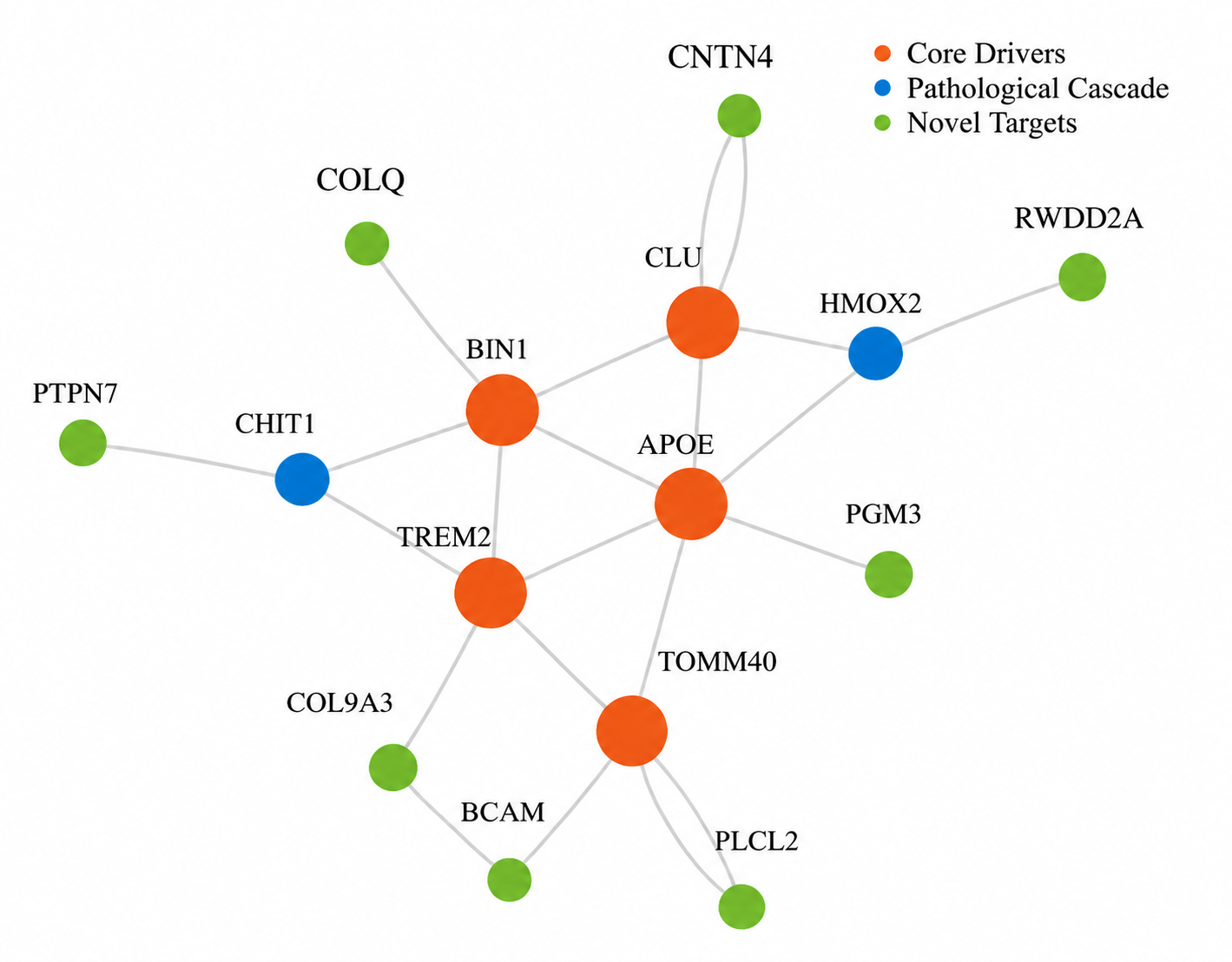}
\caption{Multi-omics integrated network of the 15 prioritized genes, categorized into Canonical Risk Loci, Secondary Response Loci, and Candidate Biomarkers.}
\label{fig_8}
\end{figure}

To examine the multimodal fusion feature learning capability of STGSCCA from a broader systems biology perspective, we took the union of the top 10 features prioritized by the model under the SNP and DNA methylation modalities, mapping them to construct a multi-omics joint gene set containing 15 core targets, as illustrated in Figure~\ref{fig_8}. This network delineates a coordinated genetic--epigenetic cooperative architecture associated with disease progression along the Alzheimer's disease (AD) continuum in EMCI, which can be categorized into three primary functional groups. The first group comprises canonical susceptibility loci, including \textit{APOE}, \textit{TOMM40}, \textit{TREM2}, \textit{BIN1}, and \textit{CLU}. As well-established genetic determinants associated with cognitive impairment across the AD continuum, these genes are centrally involved in lipid metabolism and amyloid clearance pathways that characterize early preclinical stages, demonstrating the robust signal-anchoring capacity of STGSCCA when analyzing noisy longitudinal EMCI cohorts. The second group represents secondary pathological response markers, exemplified by the neuroinflammatory indicator \textit{CHIT1} and the oxidative stress regulator \textit{HMOX2} \cite{ref47,ref48}, indicating that the cross-modal features captured by STGSCCA broadly encompass diverse dimensions of immune and metabolic dysregulation emerging during the EMCI phase. The third group consists of exploratory candidate targets, including \textit{BCAM} \cite{ref49}, \textit{CNTN4} \cite{ref50}, \textit{COL9A3}, \textit{COLQ}, \textit{PGM3}, \textit{PLCL2}, \textit{PTPN7} \cite{ref51}, and \textit{RWDD2A} \cite{ref52}. Although such peripheral targets are frequently overlooked in conventional single-modality studies due to modest univariate effect sizes, they are successfully identified here through their coordinated covariation across the SNP--Meth--neuroimaging axis, facilitated by spatiotemporal feature learning and dual-graph Laplacian regularization. Identifying this cross-modal gene set not only underscores the utility of the proposed framework in integrating multidimensional omics layers to capture EMCI pathological heterogeneity, but also provides empirical computational candidates for mapping multi-level regulatory networks and informing early therapeutic exploration during the prodromal window.

\subsubsection{Longitudinal Association Consistency of Top Imaging Genetic Traits}
\begin{figure}[htbp]
\centering
\includegraphics[width=\textwidth]{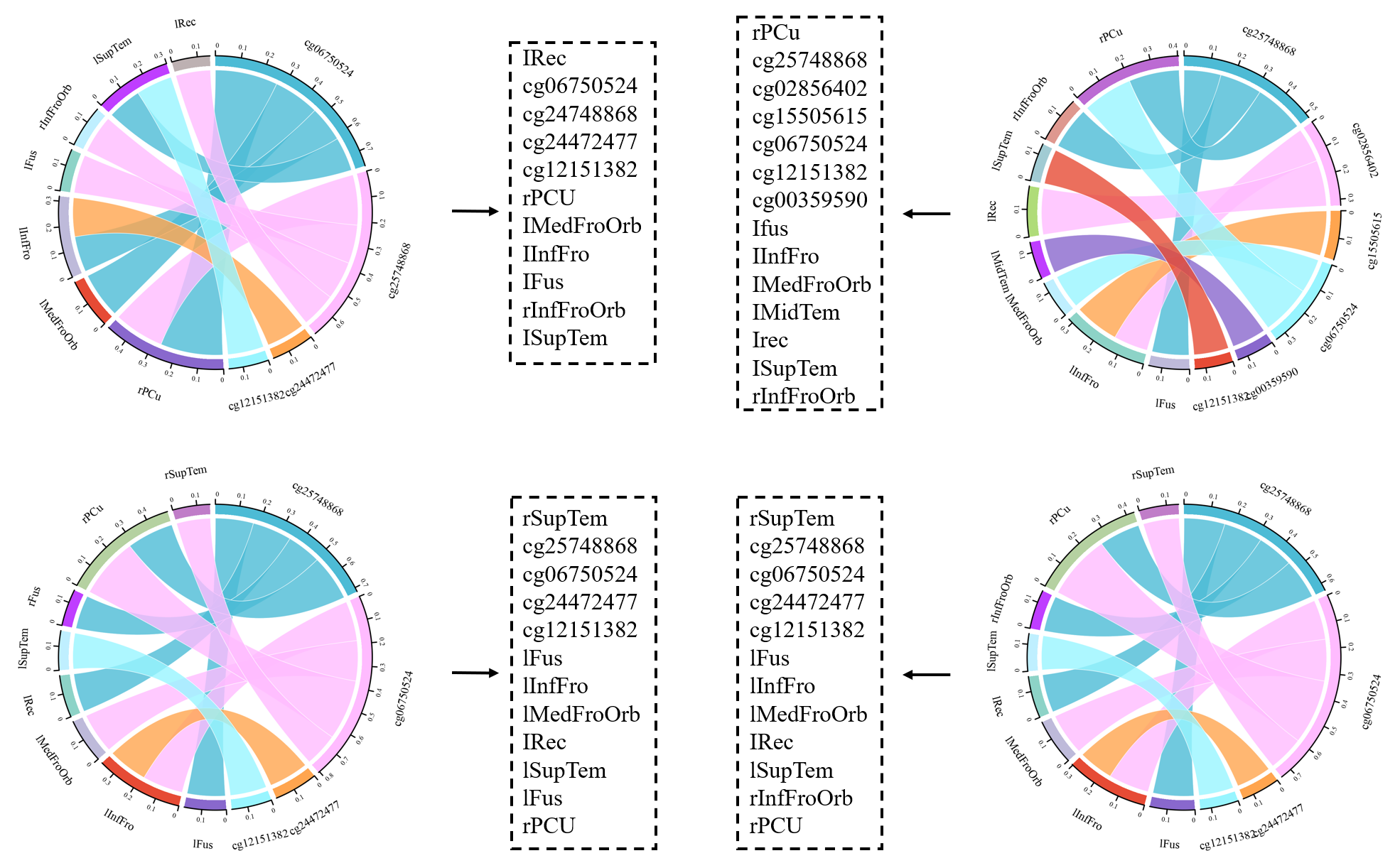}
\caption{Longitudinal partial correlation mapping of high-weight Meth-MRI association pairs at \(T_{1}\)--\(T_{4}\).}
\label{fig_9}
\end{figure}

\begin{figure}[htbp]
\centering
\includegraphics[width=\textwidth]{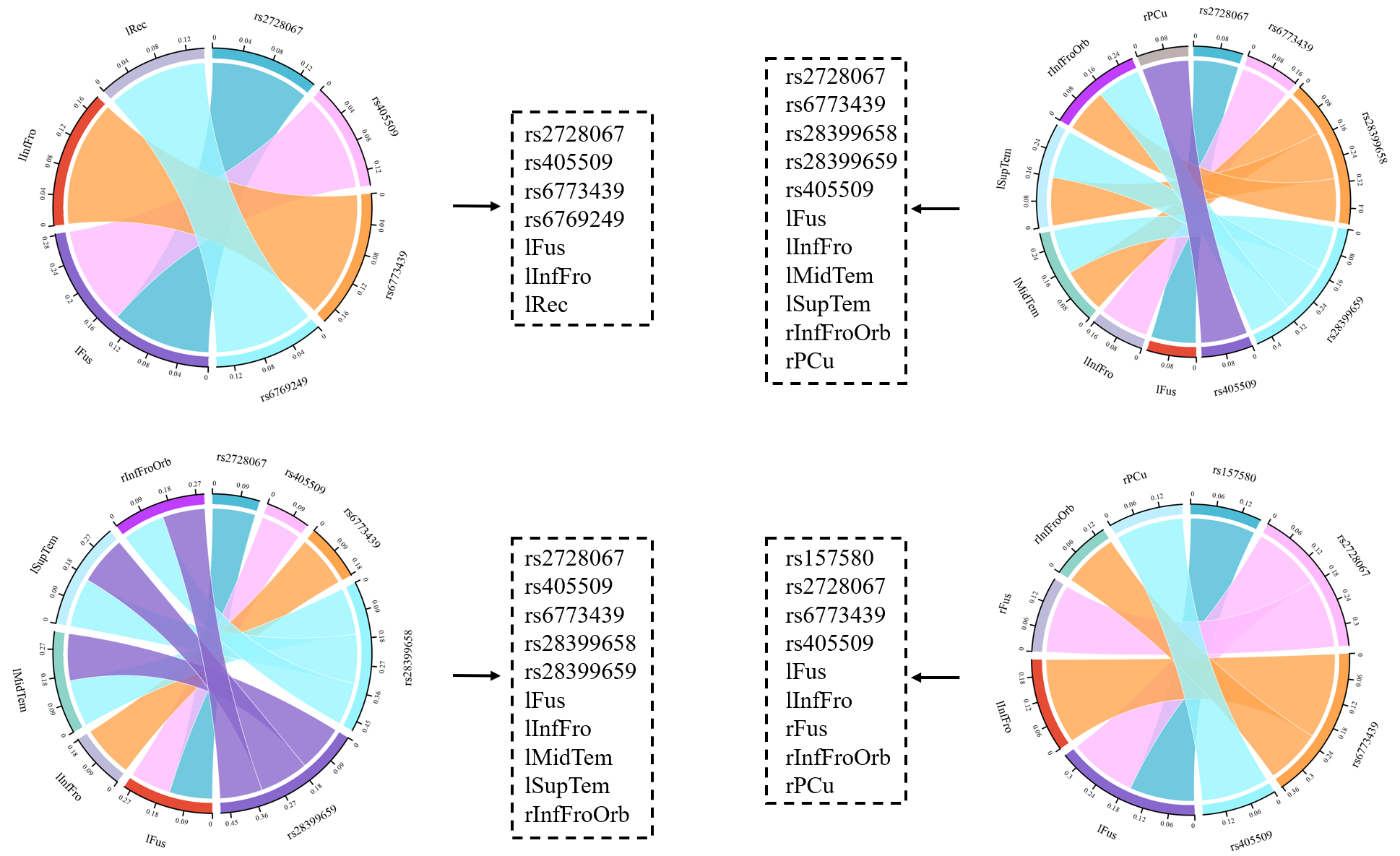}
\caption{Longitudinal partial correlation mapping of high-weight SNP-MRI association pairs at \(T_{1}\)--\(T_{4}\).}
\label{fig_10}
\end{figure}

\begin{figure}[htbp]
\centering
\includegraphics[width=\textwidth]{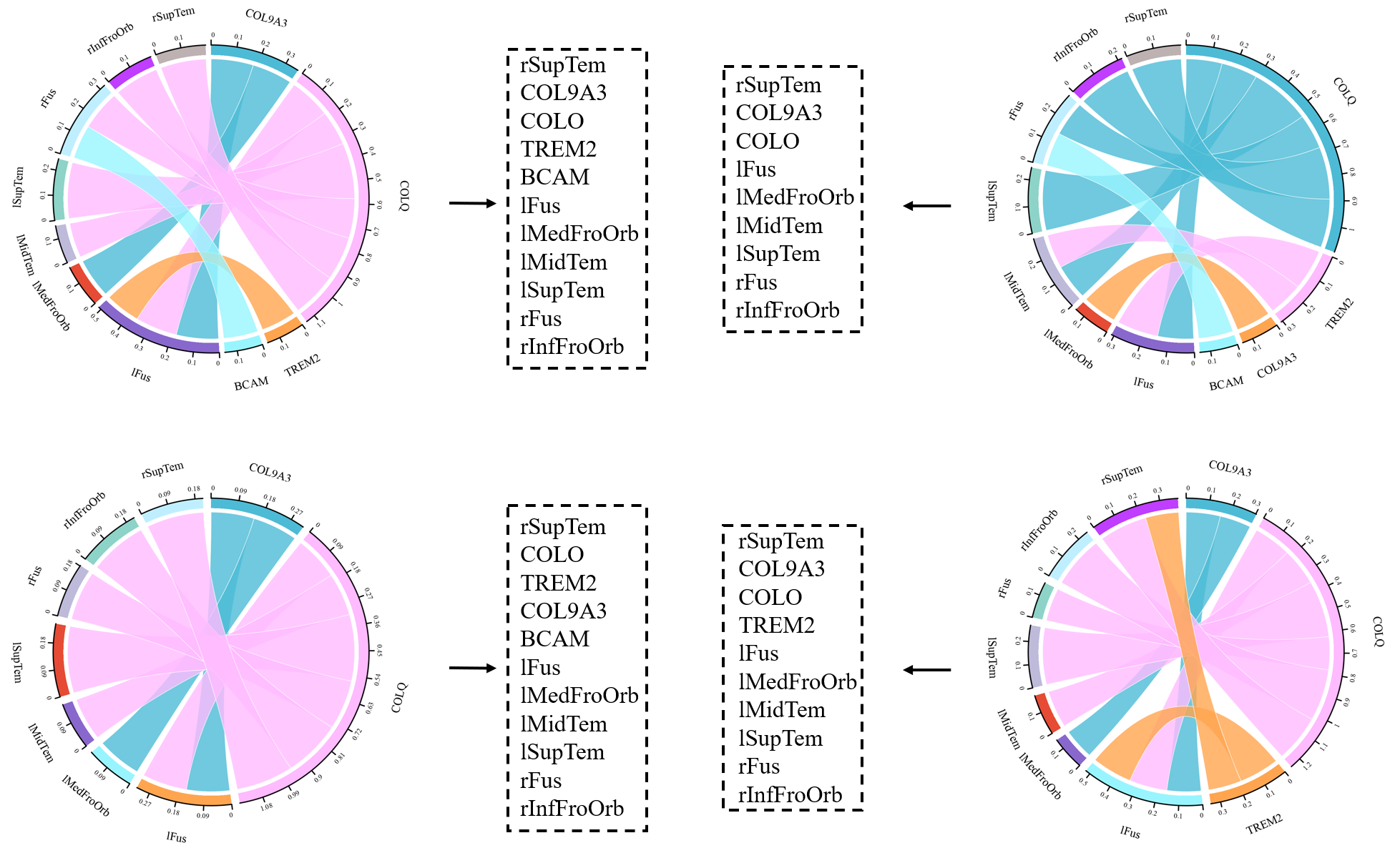}
\caption{Longitudinal partial correlation mapping of the 15 union gene sets with MRI traits at \(T_{1}\)--\(T_{4}\).}
\label{fig_11}
\end{figure}

To evaluate the biological interpretability and longitudinal stability of the features identified by STGSCCA, we examined the cross-period associations of the top-weighted multimodal targets. Specifically, we analyzed the pairwise relationships between high-weight molecular markers (SNP, DNA methylation, and the 15 joint union genes) and corresponding regions of interest (ROIs) across four consecutive follow-up time points ($T_1$ to $T_4$). As depicted in Figures~\ref{fig_9}--\ref{fig_11}, longitudinal partial correlation analyses were conducted while controlling for potential demographic confounders, including age and sex, with false discovery rate (FDR) correction applied to adjust for multiple comparisons ($P_{\text{FDR}} < 0.05$). This analysis aimed to determine whether the identified cross-modal associations exhibit persistent and reproducible patterns throughout the disease course rather than transient, single-timepoint cross-sectional artifacts.

The partial correlation trajectories demonstrate strong spatiotemporal consistency across the progression stages. At the epigenetic level, the core methylation probe cg06750524 (\textit{APOE}) and the right precuneus (rPCu) maintained significant negative associations from $T_1$ to $T_4$. Given that the precuneus represents a central hub of the DMN known to undergo early metabolic and structural vulnerability, this persistent cross-period coupling indicates that the model successfully captures epigenetic alterations stably linked to regional cortical atrophy. In the genetic dimension, the risk locus rs6773439 consistently correlated with atrophy in the left inferior frontal gyrus (lInfFro). At the multi-omics gene set level, coordinated network patterns were observed: the hub gene \textit{COLQ} exhibited sustained associations with widespread frontotemporal structures, including the superior temporal gyrus (lSupTem/rSupTem) and the orbital inferior frontal gyrus (rInfFroOrb). Notably, both the neuroinflammatory marker \textit{TREM2} (cg25748868) and the canonical \textit{APOE} variant rs405509 converged on the fusiform gyrus (lFus). This spatial convergence of top-tier risk markers across distinct modalities underscores shared neurodegenerative involvement. Overall, the temporal consistency and cross-period alignment across multiple follow-ups confirm that the captured imaging-genomic relationships reflect persistent pathological co-variation along the disease continuum, providing robust computational candidates for unraveling multi-omics mechanisms underlying brain network decline. It should be noted that these partial correlation trajectories reflect reproducible statistical co-variation rather than direct biological causality, serving to prioritize stable candidate markers for subsequent mechanistic validation.

\subsubsection{Analysis of Early Clinical Diagnostic Performance of Multimodal Fusion Features for EMCI vs CN}
To evaluate the effectiveness and discriminative power of the multimodal fusion features extracted by the proposed framework in early clinical diagnosis, this study conducted classification prediction experiments targeting EMCI and CN populations. Since EMCI belongs to the very early stage of Alzheimer's disease, its imaging and omics pathological changes are often extremely subtle, so accurate classification at this stage has long been regarded as a major challenge in the field. The experimental results clearly show that downstream conventional machine learning classifiers trained on features extracted by our core model are overall superior in comprehensive performance to existing multimodal fusion baseline algorithms such as MAGNET \cite{ref53}, ADCCA \cite{ref54}, SOM2LM \cite{ref55}, and UNCOT \cite{ref56}. The results are shown in Table \ref{tab_5}.

Specifically, our voting ensemble model demonstrated the most outstanding comprehensive diagnostic capability, achieving the highest F1 score (\(0.8140\pm 0.0592\)) and AUC value (\(0.8080\pm 0.0480\)) overall. In clinical diagnostic tasks, AUC and F1 score are core metrics for measuring the overall robustness of a model under positive-negative sample imbalance and the costs of false positives and false negatives. In addition, a simple logistic regression model also achieved the highest accuracy (\(0.7692\pm 0.0647\)) in our feature space.

In comparison, although some baseline models performed prominently on a single extreme value, for example, UNCOT had the highest recall (0.8789) but its precision dropped to 0.6305, and MAGNET had a precision as high as 0.8852 but its recall was only 0.5725. This severe metric imbalance indicates that baseline algorithms are highly prone to decision bias in complex high-noise multimodal data, causing their final F1 and AUC comprehensive scores to lag substantially behind our scheme.

The fact that conventional machine learning classifiers, such as SVM, random forest, and ensemble learning, can comprehensively surpass complex baseline algorithms on our feature set has important validation significance. It strongly proves that our core feature extraction framework successfully discards redundant clinical noise and precisely locks onto the most disease-specific and highly discriminative multimodal biomarkers. High-quality feature representations enable even structurally simple linear or tree models downstream to achieve more stable and balanced early precise screening of EMCI.

\begin{table}[htbp]
\centering
\caption{Classification evaluation results}
\label{tab_5}
\resizebox{\textwidth}{!}{%
\begin{tabular}{llccccc}
\toprule
Model & & Accuracy & Recall & Precision & F1 score & AUC \\
\midrule
MAGNET  & & \(0.6869\pm 0.0700\) & \(0.5725\pm 0.0886\) & \(\mathbf{0.8852\pm 0.0807}\) & \(0.6914\pm 0.0737\) & \(0.8039\pm 0.0542\) \\
ADCCA   & & \(0.7195\pm 0.1199\) & \(0.7041\pm 0.1169\) & \(0.8291\pm 0.1276\) & \(0.7560\pm 0.1024\) & \(0.7255\pm 0.1298\) \\
SOM2LM  & & \(0.5922\pm 0.0712\) & \(0.7053\pm 0.1053\) & \(0.6613\pm 0.0536\) & \(0.6791\pm 0.0655\) & \(0.6519\pm 0.0955\) \\
UNCOT   & & \(0.6055\pm 0.0397\) & \(\mathbf{0.8789\pm 0.0420}\) & \(0.6305\pm 0.0280\) & \(0.7338\pm 0.0276\) & \(0.7470\pm 0.0947\) \\
\midrule
\multirow{7}{*}{ours}
& Linear\_SVM            & \(0.7487\pm 0.0650\) & \(0.7246\pm 0.0803\) & \(0.8500\pm 0.0690\) & \(0.7794\pm 0.0622\) & \(0.7917\pm 0.0738\) \\
& RBF\_SVM               & \(0.7490\pm 0.0641\) & \(0.7596\pm 0.0975\) & \(0.8349\pm 0.0834\) & \(0.7878\pm 0.0541\) & \(0.8065\pm 0.0483\) \\
& Logistic\_Regression   & \(\mathbf{0.7692\pm 0.0647}\) & \(0.7480\pm 0.0943\) & \(0.8652\pm 0.0597\) & \(0.7980\pm 0.0620\) & \(0.8062\pm 0.0526\) \\
& Random\_Forest         & \(0.7490\pm 0.0561\) & \(0.8462\pm 0.0216\) & \(0.7773\pm 0.0722\) & \(0.8081\pm 0.0362\) & \(0.7773\pm 0.0594\) \\
& Gradient\_Boosting     & \(0.7216\pm 0.0938\) & \(0.8351\pm 0.0353\) & \(0.7548\pm 0.1010\) & \(0.7905\pm 0.0658\) & \(0.7346\pm 0.0574\) \\
& MLP\_NeuralNet         & \(0.7076\pm 0.0552\) & \(0.7485\pm 0.0844\) & \(0.7849\pm 0.0803\) & \(0.7597\pm 0.0420\) & \(0.7498\pm 0.0769\) \\
& Voting\_Ensemble       & \(0.7630\pm 0.0854\) & \(0.8246\pm 0.0393\) & \(0.8095\pm 0.0971\) & \(\mathbf{0.8140\pm 0.0592}\) & \(\mathbf{0.8080\pm 0.0480}\) \\
\bottomrule
\end{tabular}%
}
\end{table}

\section{Ablation Study}
\subsection{Association Based on Brain Imaging and Genetic Data}
To thoroughly evaluate the specific contribution of each module in the two-stage framework of this study, we designed an ablation experiment containing five progressive variants and used the real imaging genetics data from ADNI. The first four variants mainly focused on the evolution of prior regularization terms in the traditional mathematical objective function. The baseline model adopted only standard sparse canonical correlation analysis. Subsequently, graph Laplacian regularization (Baseline + Lap) was introduced separately to preserve the network topology of genetic features, or multi-task cooperative penalty (Baseline + Multi-task) was introduced separately to force different genetic modalities to select consistent imaging brain regions. The fourth group (Baseline + Multi-task + Lap) superimposed all constraints and represents the complete form of the traditional DG-MTSCCA back-end model. It should be particularly clarified that, since traditional SCCA algorithms are architecturally unable to directly establish high-order tensor mappings across the temporal dimension, for these four variants lacking the front-end GCL module, we degraded the model to ``independent time point analysis,'' that is, separately computing the canonical correlation between the genetic modality and the imaging features at each of \(T_{1}\) to \(T_{4}\), and taking the arithmetic mean of these 4 independent time points as the final CCC result. In contrast, the complete framework of this study (Ours) introduces the ST-GCL front-end module, which uses a dynamic temporal self-attention mechanism to losslessly compress and fuse the nonlinear imaging sequences at 4 time points into a unified high-quality spatiotemporal representation matrix in advance, which is then input as a whole into the back end for solving.

The ablation results reveal gradual performance changes and illustrate the merits of the longitudinal spatiotemporal fusion mechanism relative to cross-sectional slice-wise analysis. As shown in Table~\ref{tab_6}, among the first four model variants, introducing the multi-task cooperative term yields moderate performance gains and pushes the purely traditional combined model toward the performance limit achievable by independent single-time-point analysis, with a Mean CCC of 0.2325. However, this strategy, which computes features separately for each time point and averages outputs, carries an inherent limitation from a biological perspective: it breaks the continuity of the Alzheimer’s disease neurodegenerative trajectory, fails to capture the accelerating dynamics of regional brain atrophy, and cannot effectively suppress local high-frequency noise present in individual scans. Our full framework achieves further improvement under this setting, reaching a Mean CCC of 0.2353. This observation supports the contribution of the ST-GCL module. Instead of merely fitting static gene data to a snapshot of a single static brain network, the complete model learns the high-dimensional mapping between static genotypes and dynamic phenotypic deterioration trajectories. This enables more precise and robust identification of potential pathogenic targets across the full longitudinal disease progression.

\begin{table}[htbp]
\centering
\caption{CCCs results of Ablation Variants}
\label{tab_6}
\begin{tabular}{lccc}
\toprule
Ablation Variant & SNP CCC & DNAm CCC & Mean CCC \\
\midrule
Baseline                        & 0.1900 & 0.2460 & 0.2180 \\
Baseline + Lap                  & 0.1895 & 0.2465 & 0.2180 \\
Baseline + Multi-task           & 0.2072 & 0.2573 & 0.2323 \\
Baseline + Multi-task + Lap     & 0.2075 & 0.2574 & 0.2325 \\
Full Framework with GCL         & \(\mathbf{0.2105}\) & \(\mathbf{0.2601}\) & \(\mathbf{0.2353}\) \\
\bottomrule
\end{tabular}
\end{table}

\subsection{Selection Based on Top Genetic Features}

\begin{figure*}[htbp]
\centering
\begin{tabular}{c@{\hspace{6pt}}c}
\textbf{(a)} & \includegraphics[width=1\textwidth]{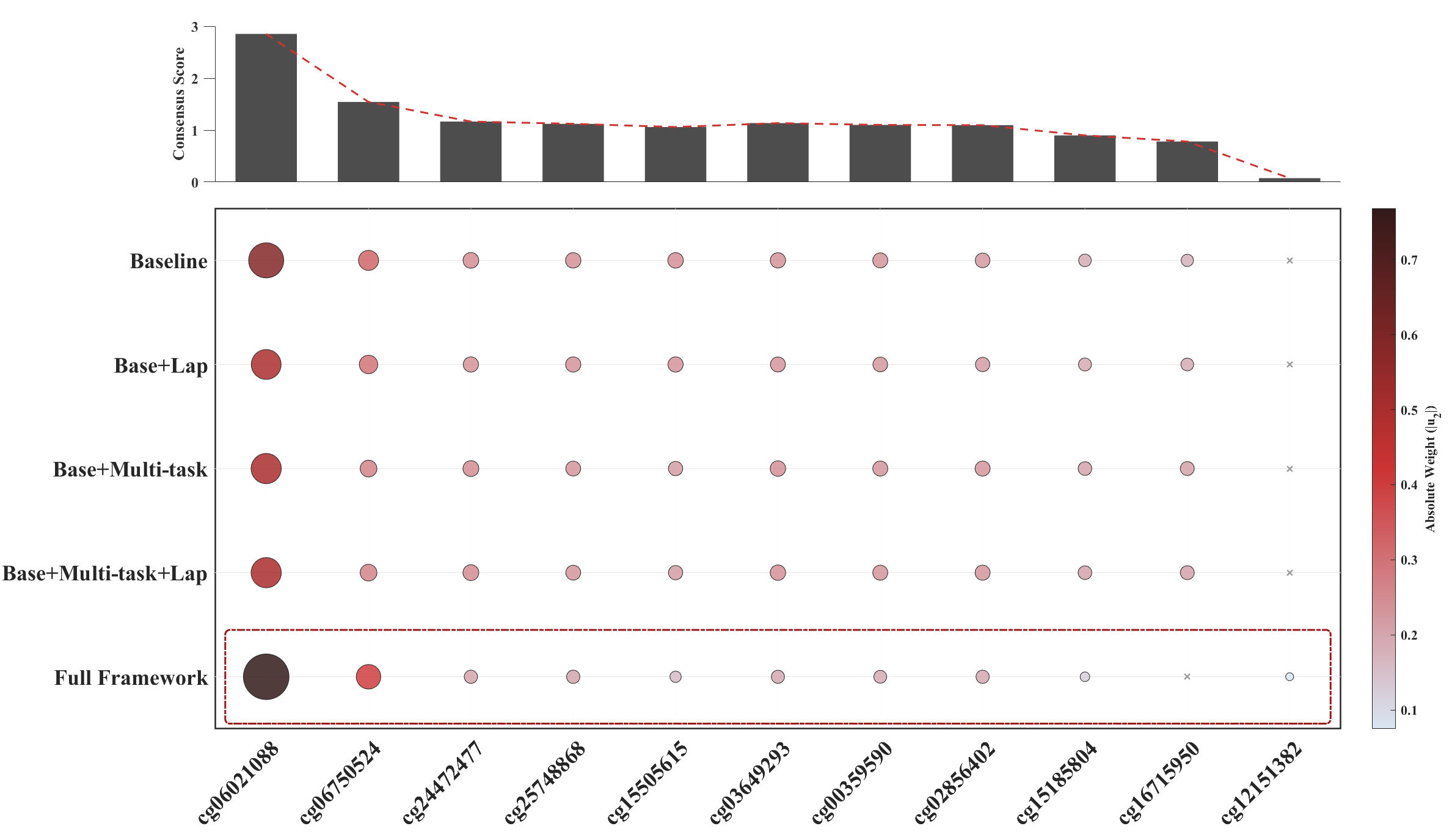} \\[6pt]
\textbf{(b)} & \includegraphics[width=1\textwidth]{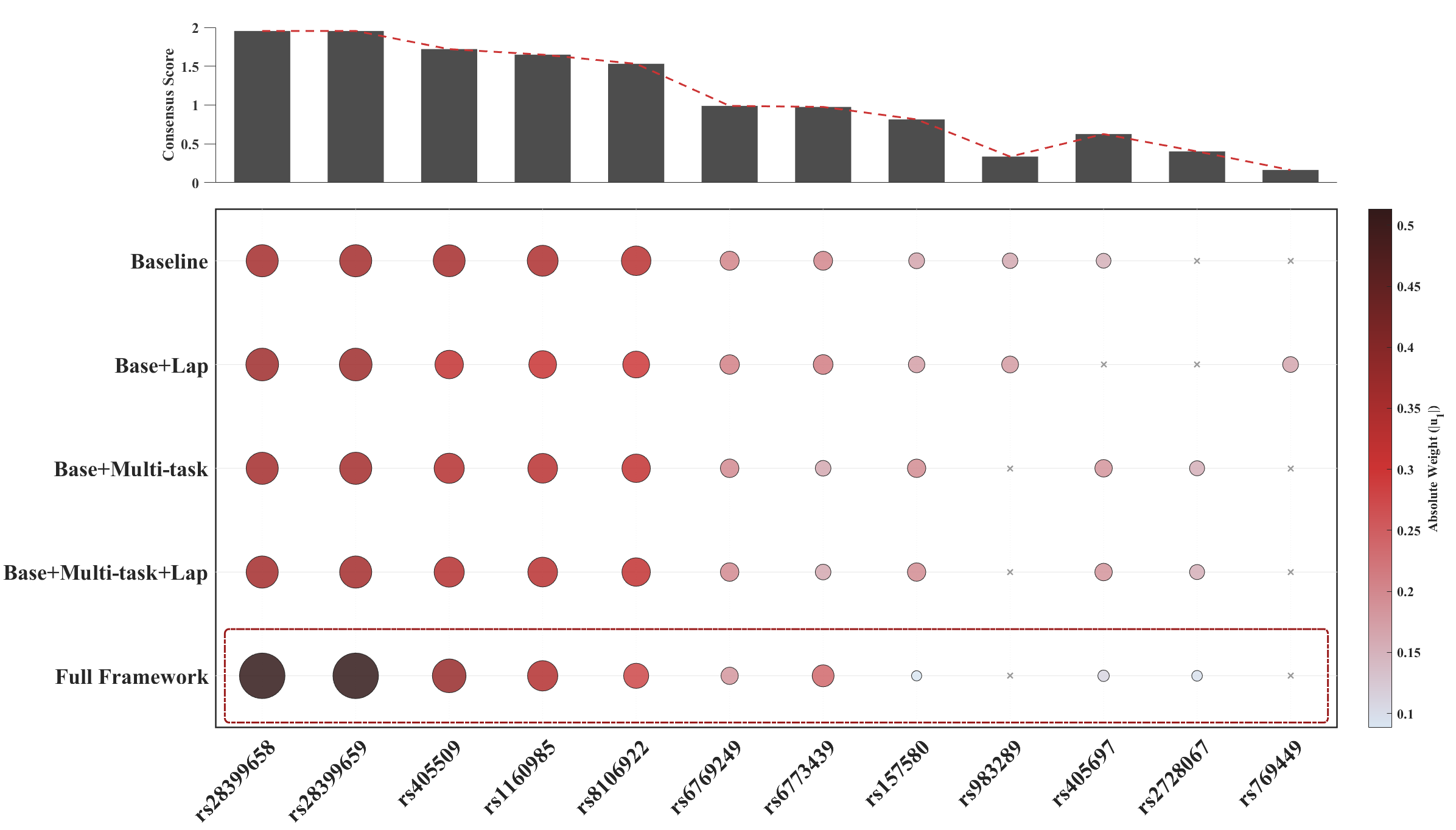} \\
\end{tabular}
\caption{Selection of top features across ablation variants for (a) top CpG loci and (b) top SNPs. The upper bar plots display the consensus scores of key loci across all variants, with red dashed lines indicating the score decay trends. The lower bubble matrices illustrate the absolute feature weights ($|\mathbf{u}_c|$) across five ablation models, where circle size and color depth represent weight magnitude, and grey crosses ($\times$) denote unselected features (zero weight). The proposed full framework is highlighted with red dashed boxes, demonstrating enhanced sparsity and polarized feature selection on primary pathogenic targets compared to the baseline variants.}
\label{fig_12}
\end{figure*}

As shown in Figure~\ref{fig_12}, the experimental results show that the baseline model without the spatiotemporal dynamic fusion module and its transitional variants exhibit relatively dispersed and flat feature weight distributions. This indicates that modeling based on independent single-time-point slices makes it difficult to suppress temporal measurement noise, resulting in insufficient focusing capability in cross-modal feature selection. In contrast, after introducing the ST-GCL front-end module, the feature weights of the complete STGSCCA framework exhibit highly sparse and focused distribution characteristics. At the SNP level, the model not only effectively penetrates complex genetic background noise, but also precisely amplifies the feature weights of recognized core pathogenic mutations of Alzheimer's disease, such as rs405509 strongly associated with \textit{APOE} and the highly linked rs28399658/rs28399659. At the DNAm level, the model achieves a precipitous signal enhancement for major-effect methylation sites such as cg06021088, while imposing severe sparse suppression on marginal covariates. The above results indicate that the spatiotemporal purification of longitudinal evolution trajectories by the front-end ST-GCL significantly suppresses omics and imaging noise, enabling the back-end DG-MTSCCA to localize key pathogenic biomarkers with a higher signal-to-noise ratio and feature specificity.

\section{Discussion}
Addressing the challenges of high-frequency measurement noise and nonlinear temporal modeling faced by longitudinal imaging genetics in the analysis of EMCI, this study achieves an innovative integration at the methodological architecture level. We propose a two-stage framework that integrates front-end deep spatiotemporal purification with back-end network association target hunting. For the first time, ST-GCL is introduced as the front-end engine to losslessly compress and purify multi-time-point longitudinal imaging sequences into a single individual-level spatiotemporal representation. Furthermore, for the first time, this deep learning-purified dynamic representation is input into a dual graph-regularized multi-task analysis (DG-MTSCCA) back end for cross-modal multi-task joint target hunting with underlying SNPs and DNA methylation. This new paradigm effectively breaks through the linear constraint bottleneck of traditional advanced longitudinal algorithms when handling complex high-dimensional temporal data.

To verify the underlying performance of the proposed algorithm within strict and controllable theoretical boundaries, this study first conducted simulated data experiments based on a preset ground truth. These experiments aimed to investigate the anti-overfitting and feature mining capabilities of the algorithm under extreme sandbox scenarios such as extremely small samples, harsh signal-to-noise ratios, and strong modular collinearity interference. The experimental results show that, driven by the powerful spatial smoothing and temporal denoising engine of the front-end ST-GCL, this model effectively filters out the random noise that causes precipitous performance decline in traditional advanced longitudinal algorithms. Meanwhile, the back-end dual graph Laplacian regularization successfully preserves the complete gene network with close topological associations, exhibiting advantageous F1 scores and AUC metrics in all test scenarios and establishing its absolute advantage in high-dimensional small-sample environments.

On the basis of theoretical validation, this study then conducted multimodal cross-omics association and feature selection experiments on the real EMCI longitudinal cohort of ADNI to evaluate the clinical interpretability and knowledge discovery capability of the model. The association results show that this framework breaks through the performance ceiling of existing advanced longitudinal algorithms in both Meth-MRI and SNP-MRI tasks. In the mining of pathogenic targets, this framework exhibits high feature selection purity. It  precisely anchors top-tier immune and lipid metabolism risk genes centered on \textit{APOE}, \textit{BIN1}, and \textit{TREM2}, and maps macroscopic phenotypes to vulnerable regions within the default mode network, such as the precuneus. Furthermore, within the joint gene set, the algorithm prioritizes several exploratory candidate biomarkers, such as \textit{COLQ} and \textit{PTPN7}, which exhibit modest univariate effects in single-modality analyses but show coordinated covariation across omics and imaging layers.

To further evaluate the longitudinal stability of the prioritized features, we conducted a cross-period partial correlation analysis across the four follow-up time points ($T_1$ to $T_4$) while controlling for age and sex. The primary objective of this analysis was to examine whether the molecular candidates identified by STGSCCA maintain reproducible statistical associations with regional brain volumes throughout the EMCI follow-up period. We observed that key marker--ROI pairs exhibit consistent cross-period correlations across multiple time points. Rather than reflecting transient single-timepoint statistical fluctuations, this spatiotemporal consistency supports the reproducibility of the captured imaging--omics covariation patterns along the disease continuum, offering prioritized candidate targets for downstream hypothesis-driven research.

Based on the high-quality multi-omics fusion features extracted above, this study conducted early clinical classification experiments targeting EMCI and CN populations to directly test their clinical translational value. In this challenging task, this framework consistently outperforms existing multimodal fusion baseline models such as MAGNET and UNCOT, achieving the highest F1 score and AUC value overall. This result confirms at the application end that the front-end ST-GCL module successfully discards redundant clinical noise, endowing downstream features with strong disease discriminative power.

To systematically evaluate the synergistic gain mechanism of spatiotemporal feature purification (front end) and graph-regularized association mining (back end) in the two-stage architecture, this study conducted ablation experiments. The experiments show that, by merely progressively introducing graph Laplacian and multi-task cooperative constraints, the representation capability of traditional mathematical models when handling multi-time-point heterogeneous data remains limited. In contrast, this framework effectively decouples complex longitudinal trajectory modeling from cross-modal association alignment through front-end spatiotemporal graph contrastive learning, contributing to more consistent cross-modal alignment and sharper feature weight focusing. Feature weight distribution analysis further indicates that the complete framework, while filtering out local measurement noise, achieves highly specific focusing and sparse selection of key pathogenic features.

Although this study has achieved significant results in longitudinal imaging genetics analysis, certain limitations remain. First, due to the extremely high screening threshold for strictly matched multi-time-point, multi-omics data, including MRI, SNP, and DNA methylation, the sample size ultimately included in this study from the ADNI cohort is relatively limited, with \(N=147\). Although the algorithm demonstrates excellent small-sample anti-overfitting capability, the generalization of the model still needs to be further verified in larger-scale, multicenter independent cohorts in the future. Second, the exploratory cross-modal candidate biomarkers prioritized by our algorithm (such as \textit{COLQ} and \textit{PTPN7}) and the observed cross-timepoint partial correlations reflect observational statistical associations rather than causal biological effects; their underlying molecular regulatory mechanisms require further verification through Mendelian randomization or wet-lab functional assays in cellular and animal models. Finally, the current imaging features focus only on gray matter volume extracted from sMRI. Future research can further introduce data such as fMRI or PET to construct a more comprehensive and three-dimensional spatiotemporal pathological evolution map of Alzheimer's disease.

\section{Conclusion}
Addressing the scan measurement noise and nonlinear disease progression commonly present in follow-up brain imaging in longitudinal imaging genetics, where existing methods mostly rely on cross-sectional single-point analysis or linear temporal assumptions, as well as the limitation of failing to effectively incorporate epigenetic modifications such as DNA methylation and their molecular intrinsic topological networks into joint analysis, this study proposes a spatiotemporal graph sparse canonical correlation analysis framework, STGSCCA, built on longitudinal neuroimaging and multi-omics data. At the front end, this system performs spatial denoising and temporal attention on brain imaging sequences through spatiotemporal graph contrastive learning, effectively extracting longitudinal phenotype trajectories representing the local morphological evolution of subjects' brains. At the back end, it uses data-driven graph Laplacian regularization to preserve, without relying on external databases, the intrinsic topological structures formed by linkage disequilibrium and co-regulation of single nucleotide polymorphisms and DNA methylation, thereby achieving effective identification of deep covariation mechanisms between longitudinal imaging evolution features and high-dimensional genetic modules. Methodologically, this research expands the joint modeling capability of longitudinal imaging genetics for complex spatiotemporal sequences and gene topological structures, and also provides a general analytical paradigm for tracking the dynamic trajectories of brain structure and the mechanisms of polygenic cooperative regulation during the pathological evolution of Alzheimer's disease.

\section*{Acknowledgments}
This work was partially supported by The Hong Kong Polytechnic University Departmental Collaborative Research Fund (Project ID: P0056428).

\bibliographystyle{unsrt}  
\bibliography{references}

\end{document}